\documentclass[journal]{IEEEtran}
\usepackage{amsmath,amsfonts,amssymb}
\usepackage{algorithmic}
\usepackage{algorithm}
\usepackage{array}
\usepackage{textcomp}
\usepackage{stfloats}
\usepackage{verbatim}
\usepackage{graphicx}
\usepackage{cite}
\usepackage{bm}
\usepackage{multirow}
\usepackage{textcomp}

\usepackage{makecell}

\usepackage{cite}
\usepackage{url}

\ifCLASSINFOpdf
\else
\fi
\usepackage{xcolor}
\usepackage{xpatch}

\usepackage{caption}

\usepackage{epstopdf}
\usepackage{mathrsfs}

\usepackage{latexsym}
\usepackage{booktabs}
\usepackage{float}
\usepackage{subfigure}

\usepackage{bbm}
\usepackage[amsmath,thmmarks]{ntheorem}

\usepackage{setspace}

\begin{document}
%
\title{Multi-modal Data based Semi-Supervised Learning for Vehicle Positioning \vspace*{0pt}}


\author{{{Ouwen Huan} ,
{Yang Yang,} \emph{Member,~IEEE} , {Tao Luo,} \emph{Senior Member,~IEEE}, {Mingzhe Chen,} \emph{Senior Member,~IEEE}}\vspace*{0pt}\\
\thanks{O. Huan and T. Luo are with the Beijing Laboratory of Advanced Information Network, Beijing University of Posts and Telecommunications, Beijing, 100876, China (e-mail: \protect\url{ouwenh@bupt.edu.cn}; \protect\url{tluo@bupt.edu.cn}).}
\thanks{M. Chen is with the Department of Electrical and Computer Engineering and Institute for Data Science and Computing, University of Miami, Coral Gables, FL, 33146, USA (e-mail: mingzhe.chen@miami.edu).}
\thanks{Y. Yang is with the Beijing Key Laboratory of Network System Architecture and Convergence, School of Information and Communication Engineering, Beijing University of Posts and Telecommunications, Beijing 100876, China (e-mail: \protect\url{yangyang01@bupt.edu.cn}).}
}

\maketitle{}
\begin{abstract}
In this paper, a multi-modal data based semi-supervised learning (SSL) framework that jointly use channel state information (CSI) data and RGB images for vehicle positioning is designed. In particular, an outdoor positioning system where the vehicle locations are determined by a base station (BS) is considered. The BS equipped with several cameras can collect a large amount of unlabeled CSI data and a small number of labeled CSI data of vehicles, and the images taken by cameras. Although the collected images contain partial information of vehicles (i.e. azimuth angles of vehicles), the relationship between the unlabeled CSI data and its azimuth angle, and the distances between the BS and the vehicles captured by images are both unknown. Therefore, the images cannot be directly used as the labels of unlabeled CSI data to train a positioning model. To exploit unlabeled CSI data and images, a SSL framework that consists of a pretraining stage and a downstream training stage is proposed. In the pretraining stage, the azimuth angles obtained from the images are considered as the labels of unlabeled CSI data to pretrain the positioning model. In the downstream training stage, a small sized labeled dataset in which the accurate vehicle positions are considered as labels is used to retrain the model. Simulation results show that the proposed method can reduce the positioning error by up to 30\% compared to a baseline where the model is not pretrained.
\end{abstract}


%
\IEEEpeerreviewmaketitle

\section{Introduction}
Vehicle positioning plays an important role for future vehicle applications such as autonomous driving and traffic monitoring \cite{1}. Current global navigation satellite system (GNSS) based vehicle localization methods (e.g. global positioning system (GPS)) have significant performance loss in urban environments due to the blockage of buildings, pedestrians, and vehicles. To improve the accuracy of these vehicle positioning methods, one can study the use of RF for vehicle positioning \cite{45,46}. Compared to GNSS based positioning methods, RF based sensing methods have lower latency and can achieve higher positioning accuracy in urban areas. Meanwhile, compared to camera based algorithms \cite{Camera}, RF based methods can localize users in both line-of-sight (LoS) and non-line-of-sight (NLoS) link dominated scenarios, and are robust to severe lighting and weather conditions. However, using RF signals for vehicle positioning still faces several challenges such as high accuracy localization of high-speed moving targets, precise 3-D signal propagation environment modeling for localizing non-line-of-sight (NLoS) users, and combination with the traditional GNSS based localization methods.

Recently, a number of existing works \cite{2,3,4,5,6,7,8,9,10} have studied the use of RF data for indoor and outdoor positioning. The work in \cite{2} first presented a hybrid location image fingerprint generated with Wi-Fi and magnetic field fingerprints, and then a convolutional neural network (CNN) was employed to classify the locations of the fingerprint images. The authors in \cite{3} designed a deep learning (DL) based localization algorithm that uses both RSS and channel state information (CSI) for indoor positioning. In \cite{4}, a positioning model based on the ResNet architecture \cite{11} is designed to localize users using NLoS transmission links. However, in these works, the RF fingerprints were collected at some fixed locations, and the positioning problem was formulated as a classification problem that classified the locations of user equipments (UE). The works in \cite{5,6,7,8,9,10} formulated positioning as a coordinate regression problem. In \cite{5}, a stacked auto-encoder with an one-dimensional CNN was developed to achieve high positioning accuracy using received signal strengths (RSS). The authors in \cite{6} used the angle-delay channel power matrix as the input of a CNN to estimate user positions. In \cite{7}, the authors designed an attention-augmented residual CNN with a larger receptive field for user positioning. The authors in \cite{8} designed two effective methods to process CSI for positioning. The work in \cite{9} studied the fingerprint based positioning aided by reconfigurable intelligent surface (RIS). In this work, the authors proposed a new type of fingerprint named space-time channel response vector (STCRV), and designed a novel residual CNN to estimate the 3D positions of targets. In \cite{10}, the authors designed a CSI fingerprint based positioning system trained with CSI collected at multiple BSs. However, all of these works \cite{2,3,4,5,6,7,8,9,10} require the use of a large amount of labeled data (i.e. RF data and their corresponding positions) to train DL models, which may not be applied for the scenarios where a server cannot collect such large amount of labeled data. 

\label{sec:2}

\begin{table}\footnotesize
\newcommand{\tabincell}[2]{\begin{tabular}{@{}#1@{}}#1.3\end{tabular}}
\renewcommand\arraystretch{1.3}
\caption[table]{{List of notations}}
\centering
\begin{tabular}{|c|c|}
\hline 
\!\textbf{Notation}\! \!\!& \textbf{Description}  \\
\hline
$V_t$ & Number of vehicles served as time $t$ \\
\hline
$P$ & Number of pilot symbol vectors \\
\hline
$\boldsymbol{h}_{t,m}$ & \makecell{Channel from BS to vehicle $m$ \\ over subcarrier $k$ at time $t$}\\
\hline 
$N_\textrm{B}$ & Antenna number of ULA\\
\hline 
$N_\textrm{C}$ & Number of OFDM sub-carriers\\
\hline 
$\boldsymbol{H}_{t,m}$ & CSI matrix of vehicle $m$ at time $t$\\
\hline 
$\boldsymbol{H}_{t}$ & CSI matrices collected by the BS at time $t$\\
\hline 
$C$ & Number of cameras equipped at BS\\
\hline 
$\boldsymbol{I}_{t,c}$ & RGB image captured by camera $c$ at time $t$\\
\hline 
$\mathcal{I}_t$ & Set of RGB images captured at time $t$\\
\hline 
$W_c$ &  Width of images captured by camera $c$\\
\hline
$H_c$ & Height of images captured by camera $c$ \\
\hline
$\left(o^w,x^w,y^w,z^w\right)$& World coordinate system \\
\hline
$\left(o^i,x^i,y^i\right)$ & Image coordinate system \\
\hline
$\left(o^p,u^p,v^p\right)$ & Pixel coordinate system \\
\hline
$\left[u_m^p, v_m^p\right]^T$ & Pixel coordinate of point $m$  \\
\hline
$\phi$ & Azimuth angle of point $m$ \\
\hline
$\theta$ & Elevation angle of point $m$ \\
\hline 
$\phi_c^L$ & \makecell{Azimuth angle of the LoS \\ direction of camera $c$} \\
\hline 
$\theta_c^L$ & \makecell{Elevation angle of the LoS \\ direction of camera $c$} \\
\hline 
$\Delta{\phi}$ & Difference between $\phi$ and $\phi_c^L$\\
\hline 
$\Delta{\theta}$ & Difference between $\theta$ and $\theta_c^L$\\
\hline 
$V_t^{'}$ & \makecell{Number of vehicles detected from \\ images in $\mathcal{I}_t$} \\
\hline 
$\phi_{t,i}$ & \makecell{Azimuth angle of vehicle $i$ \\ detected from images}\\
\hline 
$\boldsymbol{q}_t$ & \makecell{Vector contains the azimuth angles of vehicles \\ captured by the images in $\mathcal{I}_t$}\\
\hline 
\end{tabular}
\end{table}

To improve the performances of models or reduce the size of labeled dataset for model training, the works in \cite{12,13,14,15,16,18,Joint_image_RF_one,Joint_image_RF_two} studied the use of multi-modal data to assist positioning or other applications. Specifically, the authors in \cite{12} employed an object detection model to localize vehicles and blockages in images, and achieve blockage prediction with a recurrent neural network (RNN). In \cite{13}, the authors leveraged image data to proactively predict dynamic link blockages and handoff for millimeter Wave (mmWave) systems. The authors in \cite{14} and \cite{15} utilized images to achieve fast and low-overhead mmWave/Terahertz (THz) beam tracking. In \cite{16}, a fusion-based deep learning framework operating on images, LiDAR data, and GPS data was designed to predict the subset of top-K optimal beam pairs. However, none of these works considered to use multi-modal unlabeled data to generate labeled data for training DL models. Therefore, they still need to collect a large amount of labeled data to train their designed DL models. The works in \cite{18,Joint_image_RF_one,Joint_image_RF_two} studied the joint use of images and wireless data for positioning. In \cite{18}, the authors introduced an image-driven representation method to represent all the received signals with a specially designed RF image. Then, this RF image is combined with an RGB image captured by the camera for positioning. However, the work in \cite{18} did not consider how to find the correct user position in the RGB image that each RF signal corresponds to since each RGB image may capture the locations of mutiple users. The work in \cite{Joint_image_RF_one} designed an unsupervised person re-identification system using both images and wireless positioning trajectories under weak scene labeling. In \cite{Joint_image_RF_two}, the authors proposed a multi-modal context propagation framework for user localization. The proposed framework contains a recurrent context propagation module that enables position information to be fused between images and wireless data, and an unsupervised multi-modal cross-domain matching scheme that utilizes the wireless trajectories to constrain the estimation of pseudo labels of images. However, the works in \cite{Joint_image_RF_one,Joint_image_RF_two} assumed that the user locations can be directly estimated from RF data, and the positioning error of RF data is a deterministic function of the communication signal-to-noise ratio (SNR), since the focuses of these works are on how to find the correct corresponding relationship between RF fingerprints and the users in images, but not on how to train an RF fingerprint based positioning model by using vehicle positions in images as the labels of RF data.

\begin{figure}[t]
\centering
\includegraphics[width=8cm]{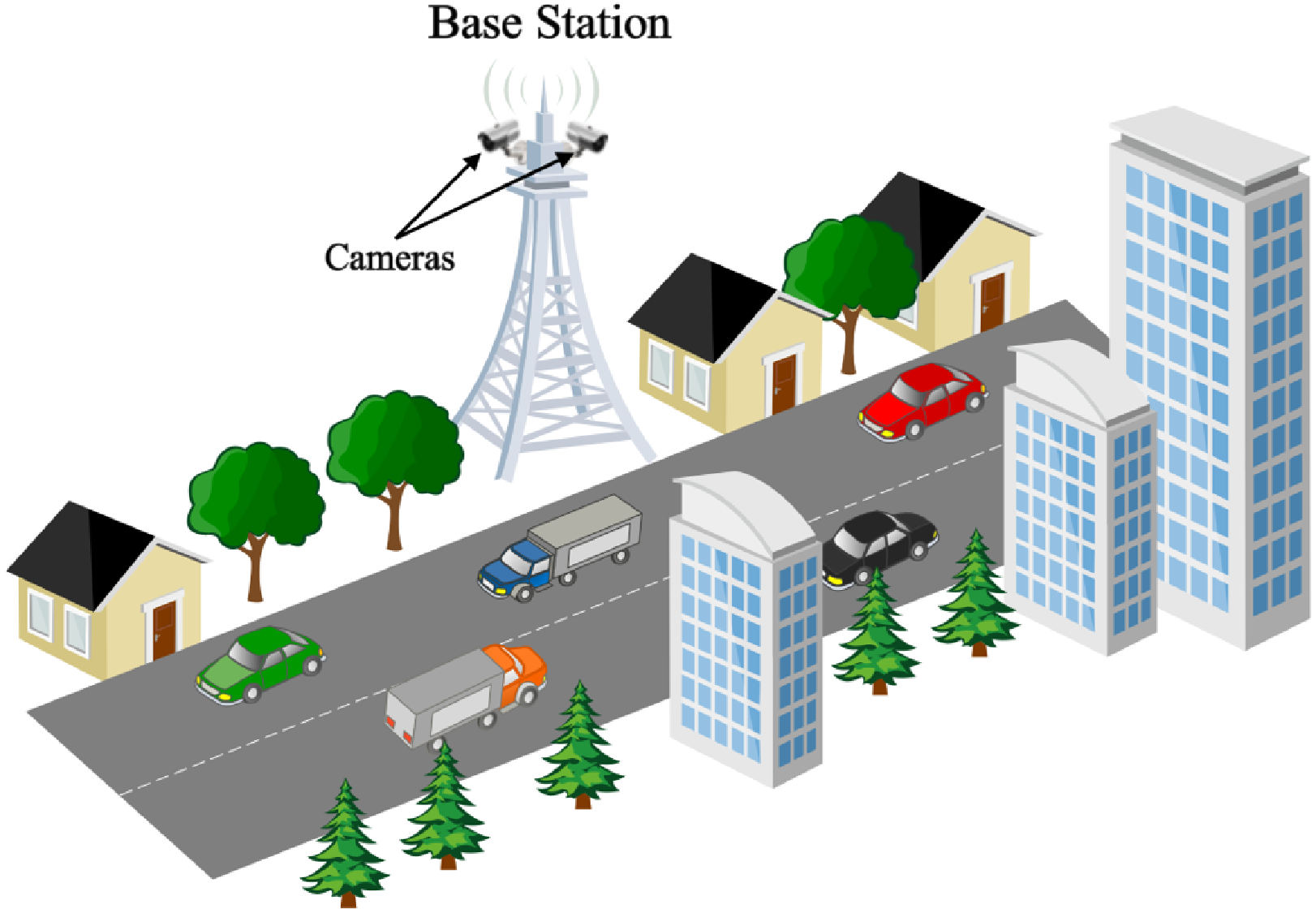}
\centering
\caption{Considered outdoor positioning scenario.}
\label{fig1}
\end{figure}

The main contribution of this work is a novel semi-supervised learning (SSL) framework that jointly uses images and unlabeled CSI data which can be collected at a lower cost to improve the performance of the vehicle positioning model in an outdoor environment. The key contributions include:
\begin{itemize}
\item We propose a novel SSL framework that jointly uses images, unlabeled data and labeled data to estimate the locations of multiple vehicles. The proposed SSL framework consists of a pretraining stage and a downstream training stage. In the pretraining stage, we consider the directional information (i.e. azimuth angles) of vehicles obtained from the images as the labels of CSI data to pretrain a part of the positioning model. Then, in the downstream training stage, a small-sized labeled dataset in which the precise vehicle positions are considered as labels is used to retrain the model.
\item Since we do not know the corresponding relationship between unlabeled CSI and vehicle azimuth angles obtained from images, we cannot directly use vehicle azimuth angles as the labels of unlabeled CSI samples for positioning model pretraining. To solve this problem, we propose to use Gaussian filter to convert the azimuth angle of each vehicle into a vector that represents the distribution of each vehicle in the angular domain. Then, using the NOR model \cite{19}, we combine all the probability distribution vectors into one vector which represents the distribution of all vehicles in the angular domain at each time slot.
\item Given the images and CSI data, we formulate the pretraining goal as predicting the probability distribution vector of all vehicles in the angular domain at each time slot. Specifically, we first use the positioning model to predict the azimuth angle of the CSI of each vehicle. Subsequently, using the NOR model, we combine the predicted probability distribution vectors into one vector which represents the predicted distribution of all vehicles in the angular domain. By minimizing the error between the predicted distribution vector and the distribution vector obtained from images, we show that the model can learn to predict the vehicle locations according to their CSI with a small-sized labeled dataset.
\end{itemize}
Simulation results show that the proposed method can significantly reduce the positioning error by up to 30\% especially when the amount of labeled data is small compared to a baseline where the model is not pretrained. To our best knowledge, this is the first work that considers joint use of CSI data and camera images for vehicle positioning. 

The rest of this paper is organized as follows. The system model of the considered outdoor vehicle positioning system which includes the radio frequency CSI collection and images acquisition and processing are described in Section $\textrm{\uppercase\expandafter{\romannumeral2}}$. Section $\textrm{\uppercase\expandafter{\romannumeral3}}$ introduces the proposed SSL framework which can jointly use images and unlabeled CSI data to improve the performance of the positioning model. In Section $\textrm{\uppercase\expandafter{\romannumeral4}}$, numerical results are presented and discussed. Finally, conclusions are drawn in Section $\textrm{\uppercase\expandafter{\romannumeral5}}$.

\section{System Model}

As shown in Fig. \ref{fig1}, we consider an outdoor millimeter wave (mmWave) positioning system where the locations of vehicles are determined with one single BS. 
{The BS is equipped with a set of $C$ cameras such that it can use both images captured by its cameras and CSI received from vehicles for vehicle positioning.} To localize vehicles, the BS will first send pilot signals to the vehicles which will send their estimated CSI information back to BS. Then, the BS can use both CSI information and images captured by their cameras to determine the locations of vehicles. Next, we will first introduce radio frequency CSI collection. Then, we introduce the images acquisition and processing in detail.

\subsection{Radio Frequency CSI Collection}
We assume that the BS is serving $V_t$ vehicles at time slot $t$. The channel between the BS and vehicle $m$ over subcarrier $k$ is defined as $\boldsymbol{h}_{t,m}^k \in\mathbb{C}^{N_\textrm{B} \times 1}$ with $N_B$ being the number of antennas of the uniform linear array (ULA) at BS. To obtain $\boldsymbol{h}_{t,m}^{k}$, the BS transmits $P$ ($P \textgreater N_\textrm{B}$) predefined pilot symbol vectors over each subcarrier to vehicle $m$ in time slot $t$. Then,
the channel $\boldsymbol{h}_{t,m}^k$ can be determined at vehicle $m$ based on the least-square (LS) criterion \cite{MIMO_estimation,channel_estimation_model}, and will be transmitted back to the BS. The received CSI matrix at BS is expressed as
\begin{equation}\label{eq:Channel Matrix}
\boldsymbol{H}_{t,m}=\left[{\boldsymbol{h}_{t,m}^1},{\boldsymbol{h}_{t,m}^2},\cdots,{\boldsymbol{h}_{t,m}^{N_{\textrm{C}}}}\right]\in\mathbb{C}^{N_\textrm{B}\times N_\textrm{C}},
\end{equation}
where $N_\textrm{C}$ is the number of valid subcarriers. We store the CSI matrices collected by the BS at current time $t$ into a three-dimensional matrix as
$
\boldsymbol{H}_{t}=\left[ \boldsymbol{H}_{t,1},\boldsymbol{H}_{t,2},\cdots,\boldsymbol{H}_{t,V_t}\right]\in\mathbb{C}^{V_t \times N_\textrm{B}\times N_\textrm{C}}.
$
We assume that most of the served vehicles cannot provide their position information to the BS when uploading their CSI, which indicates that the collected CSI will constitute a large unlabeled dataset and a small labeled dataset for training a CSI fingerprint based positioning model.

\begin{figure}[t]
\centering
\includegraphics[width=8cm]{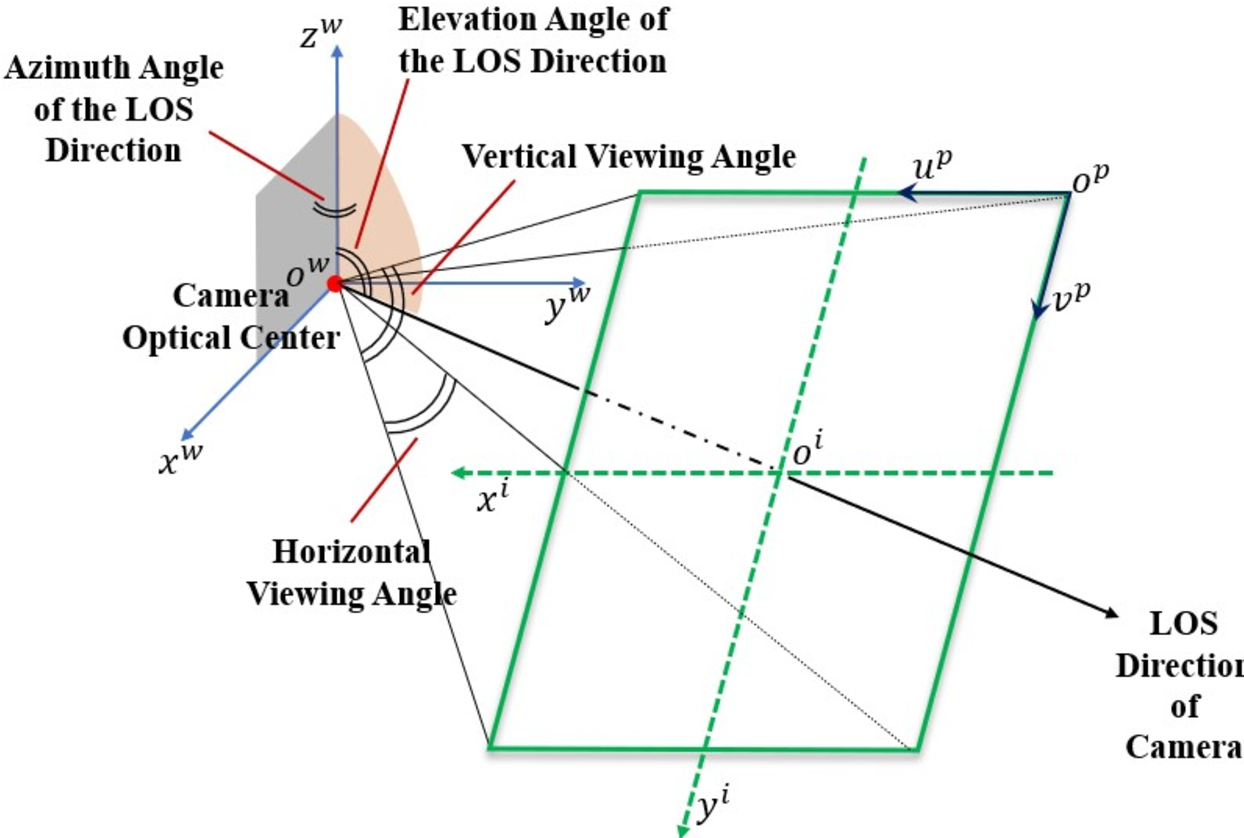}
\centering
\caption{Diagram of camera FoV.}
\label{fig2}
\end{figure}

\begin{figure}[t]
  \centering
  \subfigure[Horizontal section of camera FoV]{\includegraphics[width=7cm]{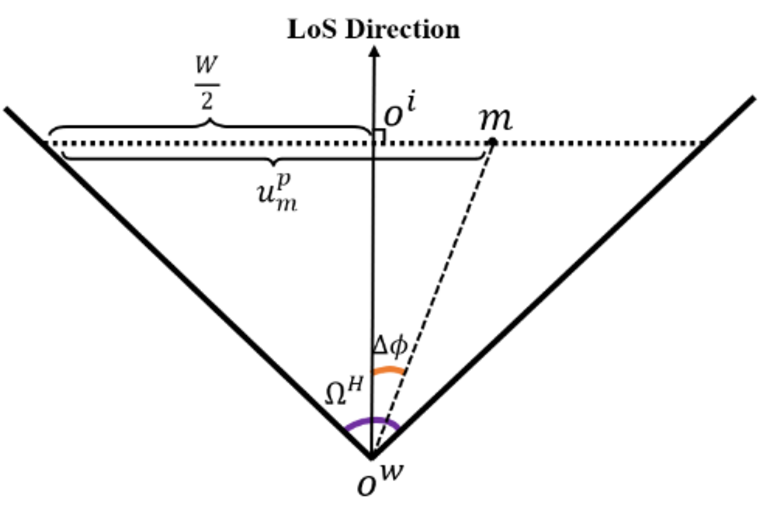}}
  \subfigure[Vertical section of camera FoV]{\includegraphics[width=6.5cm]{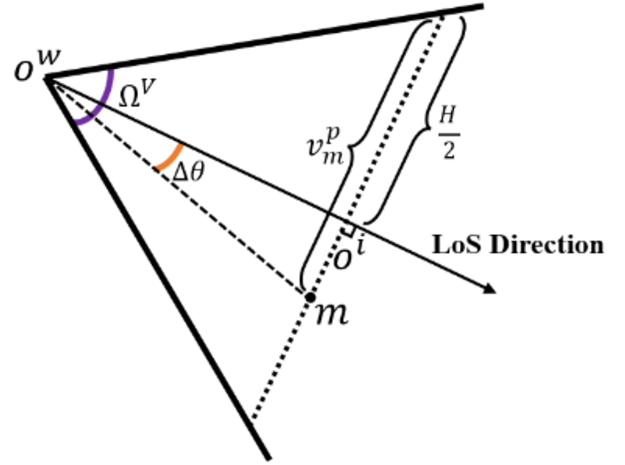}}
  \caption{Horizontal and Vertical Sections of FoV }
  \label{fig3}
\end{figure}

\subsection{Image Acquisition and Processing}
The BS is equipped with a set of $C$ cameras which will capture $C$ images at each time slot to achieve wider field of view (FoV) coverage. The set of images captured at time $t$ is
\begin{equation}\label{eq:Set of Images at time t}
\mathcal{I}_{t}=\left\{ \boldsymbol{I}_{t,c}\vert c=1,2,\cdots,C\right\},
\end{equation}
where $\boldsymbol{I}_{t,c}$ is the RGB image captured by camera $c$ at time $t$. We assume that all images taken by camera $c$ have the same dimension of $3\times W_c\times H_c$ with $W_c$ and $H_c$ respectively being the width and height of images.

To explain the use of images for vehicle positioning, we first introduce three coordinate systems shown in Fig. \ref{fig2}, which are the 3D world coordinate system (WCS) $\left(o^w,x^w,y^w,z^w\right)$, the 2D image coordinate system (ICS) $\left(o^i,x^i,y^i\right)$ and the 2D pixel coordinate system (PCS) $\left(o^p,u^p,v^p\right)$ \cite{21,22}. Both the ICS and PCS are on the image plane but have their own coordinate origins (i.e. $o^i \neq o^p$). In particular, we assume that both axis $o^{p} u^{p}$ and axis $o^{i} x^{i}$ are parallel to plane $x^{w} o^{w} y^{w}$. Given the line-of-sight (LoS) direction and viewing angles of a camera, each pixel on the image plane can be transformed to a polar coordinate $\left[\phi,\theta \right]$ in the WCS where $\phi$ and $\theta$ respectively denote the azimuth and elevation angles \cite{23}. The coordinate transformation from pixel coordinate in PCS to polar coordinate in WCS is summarized in the following lemma.

\begin{figure*}[t]
\centering
\includegraphics[width=16cm]{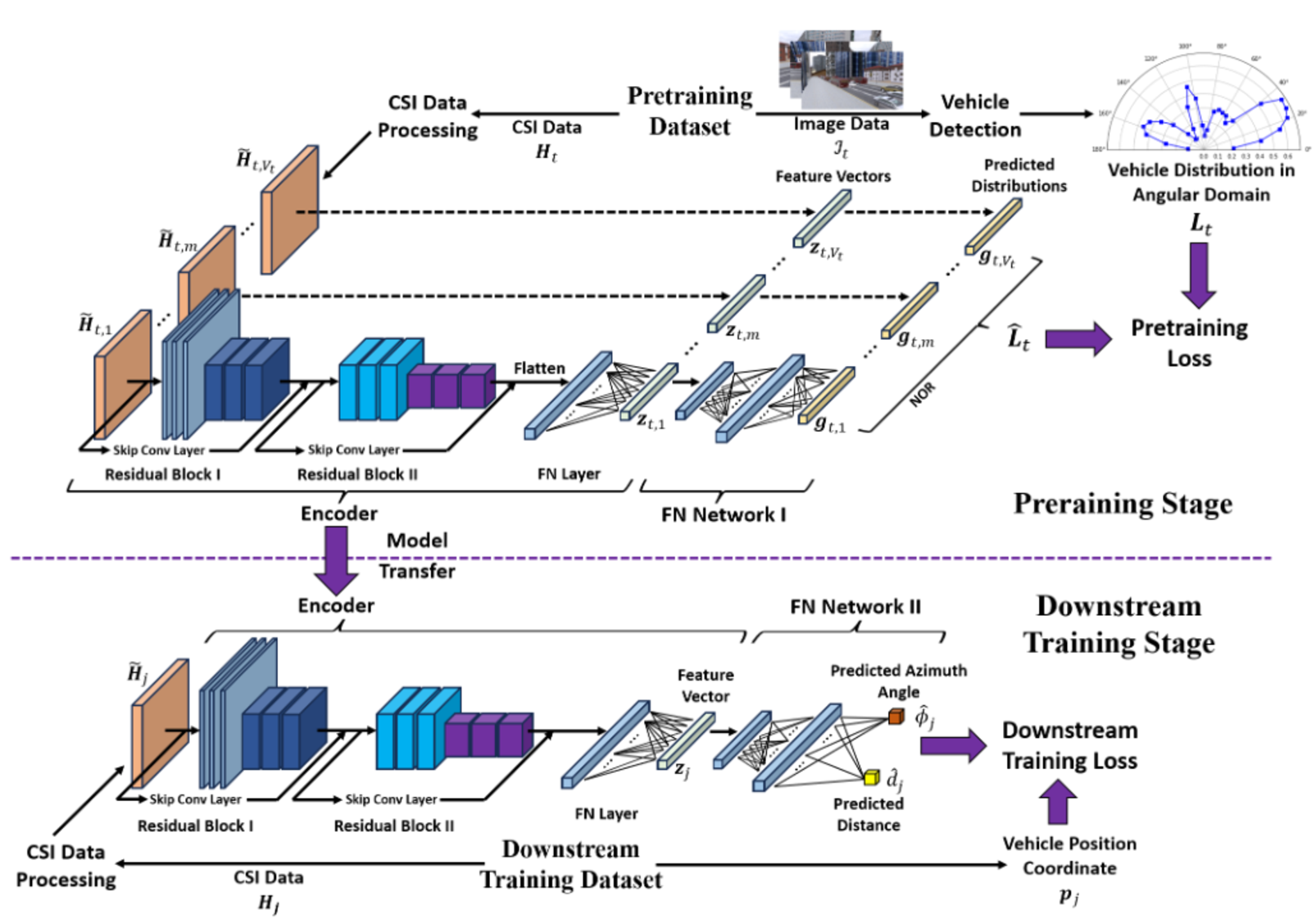}
\caption{Architecture of the proposed SSL framework.}
\label{fig4}
\end{figure*}

\itshape \textbf{Lemma 1:}  \upshape
Given the pixel coordinate of point $m$ as $\left[u_m^p, v_m^p\right]^T$, the azimuth angle $\phi$ and elevation angle $\theta$ of point $m$ in WCS are
\begin{equation}\label{eq:Set of Images at time t}
\left[ 
\begin{array}{c}
\phi \\
\theta
\end{array}
\right]
=
\left[ 
\begin{array}{c}
\phi^L \\
\theta^L
\end{array}
\right]
+
\left[ 
\begin{array}{c}
\arctan{\left(\frac{2 u_m^p - W}{W} \tan{\frac{\Omega^H}{2}}\right)} \\
\arctan{\left(\frac{2 v_m^p - H}{H} \tan{\frac{\Omega^V}{2}}\right)}
\end{array}
\right],
\end{equation}
where $\phi^L$ and $\theta^L$ respectively denote the azimuth and elevation angles of the LoS direction of the camera. $\Omega^H$ and $\Omega^V$ respectively represent the camera's horizontal and vertical viewing angles. $W$ and $H$ are the width and height of the pixel plane.

\itshape \text{Proof:}  \upshape
To determine the azimuth and elevation angles of point $m$, we first need to calculate the azimuth angle difference $\Delta{\phi}=\phi - \phi^L$ and the elevation angle difference $\Delta{\theta}=\theta - \theta^L$ between the direction of point $m$ and the camera LoS direction $\left[\phi^L,\theta^L\right]$. According to the horizontal section and vertical section of the camera FoV as shown in Fig. \ref{fig3}, we have
\begin{equation}\label{eq:Angle Interval Center}
\frac{\tan{\Delta{\phi}}}{\tan{\frac{\Omega^H}{2}}}=\frac{u_m^p-\frac{W}{2}}{\frac{W}{2}},
\end{equation}
and
\begin{equation}\label{eq:Angle Interval Center}
\frac{\tan{\Delta{\theta}}}{\tan{\frac{\Omega^V}{2}}}=\frac{v_m^p-\frac{H}{2}}{\frac{H}{2}}.
\end{equation}
Given (4) and (5), we have
\begin{equation}\label{eq:Set of Images at time t}
\left[ 
\begin{array}{c}
\tan{\Delta{\phi}} \\
\tan{\Delta{\theta}}
\end{array}
\right]
=
\left[ 
\begin{array}{c}
{\frac{u_m^p - \frac{W}{2}}{\frac{W}{2}} \tan{\frac{\Omega^H}{2}}} \\
{\frac{v_m^p - \frac{H}{2}}{\frac{H}{2}} \tan{\frac{\Omega^V}{2}}}
\end{array}
\right].
\end{equation}
Then, the azimuth angle difference and elevation angle difference are
\begin{equation}\label{eq:Set of Images at time t}
\left[ 
\begin{array}{c}
\Delta{\phi} \\
\Delta{\theta}
\end{array}
\right]
=
\left[ 
\begin{array}{c}
\arctan{\left(\frac{2 u_m^p - W}{W} \tan{\frac{\Omega^H}{2}}\right)} \\
\arctan{\left(\frac{2 v_m^p - H}{H} \tan{\frac{\Omega^V}{2}}\right)}
\end{array}
\right].
\end{equation}
Since $\phi=\phi^L + \Delta{\phi}$ and $\theta=\theta^L + \Delta{\theta}$, we have
\begin{equation}\label{eq:Set of Images at time t}
\left[ 
\begin{array}{c}
\phi \\
\theta
\end{array}
\right]
=
\left[ 
\begin{array}{c}
\phi^L \\
\theta^L
\end{array}
\right]
+
\left[ 
\begin{array}{c}
\arctan{\left(\frac{2 u_m^p - W}{W} \tan{\frac{\Omega^H}{2}}\right)} \\
\arctan{\left(\frac{2 v_m^p - H}{H} \tan{\frac{\Omega^V}{2}}\right)}
\end{array}
\right].
\end{equation}
This completes the proof.
\hfill $\Box$

Lemma 1 implies that if we can determine the pixel coordinate of a vehicle in an image, we can transform this pixel coordinate to a polar coordinate $\left[\phi,\theta \right]$ in WCS. Note that $\left[\phi,\theta \right]$ is an incomplete polar coordinate since it does not include the distance between the vehicle and camera. In other words, this polar coordinate only contains the direction information of a vehicle. The goal of positioning is to determine the horizontal polar coordinate $\boldsymbol{p}=\left[\psi,d_{\textrm{xoy}}\right]$ of each vehicle in WCS with $\psi$ and $d_{\textrm{xoy}}$ respectively being the azimuth angle and horizontal distance. Hence, given the set of captured images $\mathcal{I}_{t}$ at time $t$, we first utilize YOLOv4 \cite{24} to detect vehicles from these images and thereby obtain their horizontal pixel coordinates. Here, we can also employ other mature object detection technologies to detect vehicles in the captured images. We assume that the horizontal LoS directions and viewing angles of all cameras equipped by BS are known in advance such that we can then transform the pixel coordinate of each detected vehicle to its azimuth angle in the 3D world polar coordinate system according to lemma 1. The vector that contains the azimuth angles of the vehicles captured by the images in $\mathcal{I}_{t}$ is defined as
$\boldsymbol{q}_{t}=\left[ \phi_{t,1},\phi_{t,2}\cdots,\phi_{t,V_t^{'}}\right]\in\mathbb{R}^{1 \times V_t^{'}},$ where $V_t^{'}$ denotes the number of vehicles detected from $\mathcal{I}_{t}$. Here, $V_t^{'}$ is not always equal to $V_t$ since several vehicles may not be served by the BS or not be captured by the cameras. 





\section{Proposed Semi-Supervised Learning Framework}
\label{sec:3}
The goal of our work is to design a novel positioning framework that jointly uses a large sized unlabeled dataset that consists of images and unlabeled CSI data, and a small sized labeled dataset that consists of CSI data and their corresponding position coordinates to estimate the locations of vehicles served by the BS. Compared with the current fingerprint based positioning algorithms which heavily depend on large sized labeled training datasets, our SSL framework is trained by images and unlabeled CSI data and a small-sized labeled CSI dataset, thus achieving higher positioning accuracy. Next, we first explain the methods of processing the CSI and image data. Then, we introduce the components of the designed SSL framework. Finally, we introduce the complete model training process that consists of two stages \cite{25}:
1) pretraining with images and unlabeled CSI data, 2) downstream training with small sized labeled dataset.


\begin{figure}[t]
\centering
\includegraphics[width=9.5cm]{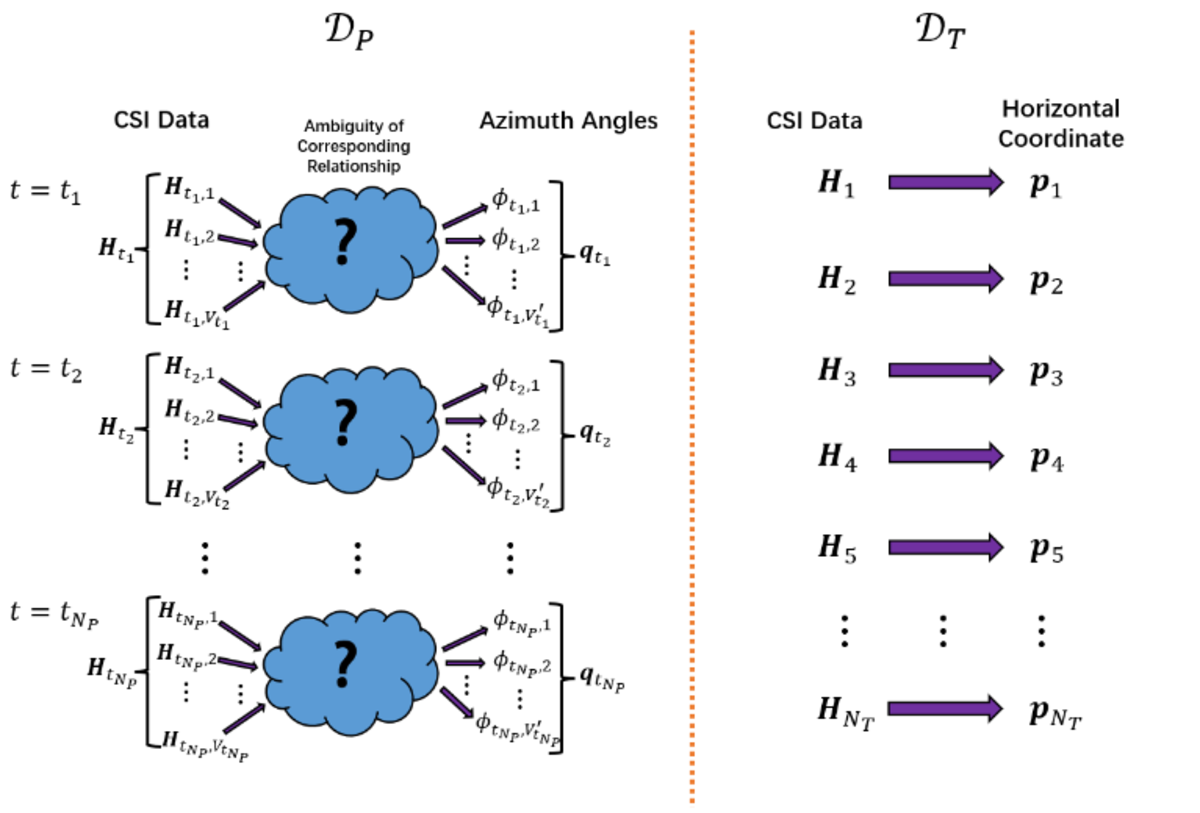}
\caption{Datasets for Pretraining and Downstream Training.}
\label{fig5}
\end{figure}

\subsection{Data Processing}
As shown in Fig. \ref{fig5}, the datasets for pretraining (i.e. images and unlabeled CSI data) and downstream training (i.e. labeled CSI data) are given by
\begin{equation}\label{eq:Pretraining Dataset}
\mathcal{D}_{\textrm{P}}=\left\{ \boldsymbol{H}_{t},\boldsymbol{q}_{t}\right\}_{t=t_1}^{t_{N_{\textrm{P}}}},
\end{equation}
\begin{equation}\label{eq:Downstream training Dataset}
\mathcal{D}_{\textrm{T}}=\left\{ \boldsymbol{H}_{j},\boldsymbol{p}_{j}\right\}_{j=1}^{N_{\textrm{T}}},
\end{equation}
where $\mathcal{D}_{\textrm{P}}$ is the pretraining dataset and $\mathcal{D}_{\textrm{T}}$ is used for downstream training. In $\mathcal{D}_{\textrm{P}}$, $t$ is the sampling time, while in downstream training dataset $\mathcal{D}_{\textrm{T}}$, $\boldsymbol{H}_{j}\in\mathbb{C}^{N_\textrm{B}\times N_\textrm{C}}$ is CSI sample $j$ and $\boldsymbol{p}_{j}=\left[x_j,y_j\right]$ denotes the corresponding position coordinate of $\boldsymbol{H}_{j}$. Next, we introduce the processing method to process CSI data $\boldsymbol{H}_{t}$ and  $\boldsymbol{H}_{j}$. Then, we explain the processing method of $\boldsymbol{q}_{t}$ for pretraining.


\subsubsection{CSI Data Processing}
We first introduce the method of processing CSI matrices of the pretraining dataset $\mathcal{D}_{\textrm{P}}$. Considering that the original CSI matrices $\boldsymbol{H}_{t,m}$ are complex-valued, we transform them to real-valued matrices. Specifically, we use the method in \cite{8} to process each complex-valued $\boldsymbol{H}_{t,m}\in{\mathbb{C}^{N_\textrm{B}\times N_\textrm{C}}}$ in $\boldsymbol{H}_{t}$ to obtain three-dimensional real-valued matrix $\widetilde{\boldsymbol{H}}_{t,m}\in\mathbb{R}^{3 \times N_\textrm{B}\times N_\textrm{C}}$ . Let $h_{t,m}^{n,k}\in\mathbb{C}^{1\times1}$ denote the element of $\boldsymbol{H}_{t,m}$ at row $n$ and column $k$. We calculate the phase difference of two adjacent elements in the same column, which is given as
\begin{equation}\label{eq:Processing CSI 1}
\delta_{t,m}^{n,k} = \arg\left(h_{t,m}^{n,k}\right) - \arg\left(h_{t,m}^{\left(n+1\right){\textrm{mod}{N_\textrm{B}}},k}\right),
\end{equation}
where $\arg\left(\cdot\right)$ denotes the argument of a complex-valued number and the mod operation implies that when $n=N_\textrm{B}$, we will calculate the phase difference between row $N_{\textrm{B}}$ and the first row. Then, one complex-valued $h_{t,m}^{n,k}$ can be transformed into a real-valued vector $\widetilde{\boldsymbol{h}}_{t,m}^{n,k}$ that consists of three elements, which is given by
\begin{equation}\label{eq:Processing CSI 1}
\widetilde{\boldsymbol{h}}_{t,m}^{n,k} = \left[\left|h_{t,m}^{n,k} \right| ,\sin{\left( {\delta_{t,m}^{n,k}}\right)},\cos{\left( {\delta_{t,m}^{n,k}}\right)}\right]\in\mathbb{R}^{1\times 3},
\end{equation}
where $\vert \cdot \vert$ denotes the modulus of a complex number, and $\widetilde{\boldsymbol{h}}_{t,m}^{n,k}$ is the three-channel vector at row $n$ and column $k$ of $\widetilde{\boldsymbol{H}}_{t,m}$. We then combine all $\widetilde{\boldsymbol{H}}_{t,m}$ into a four-dimensional matrix $\widetilde{\boldsymbol{H}}_t$ given by
\begin{equation}\label{eq:Channel Matrix}
\widetilde{\boldsymbol{H}}_t = \left[\widetilde{\boldsymbol{H}}_{t,1},\widetilde{\boldsymbol{H}}_{t,2},\cdots,\widetilde{\boldsymbol{H}}_{t,V_t} \right] \in\mathbb{R}^{V_t \times 3 \times N_\textrm{B}\times N_\textrm{C}}.
\end{equation}
where $V_t$ denotes the number of original CSI matrices in $\boldsymbol{H}_t$. 

For the downstream training dataset $\mathcal{D}_{\textrm{T}}$ and validation dataset $\mathcal{D}_{\textrm{d}}$, we respectively transform the complex-valued CSI matrices $\boldsymbol{H}_{j},\boldsymbol{H}_{j}^{\textrm{d}}$ to real-valued three-dimensional matrices $\widetilde{\boldsymbol{H}}_{j},\widetilde{\boldsymbol{H}}_{j}^{\textrm{d}}\in\mathbb{R}^{3 \times N_\textrm{B}\times N_\textrm{C}}$ using the same method for processing $\boldsymbol{H}_{t,m}$ as described in (11) and (12).
\begin{figure}[t]
\centering
\setlength{\belowcaptionskip}{-0.2cm}
\includegraphics[width=9cm]{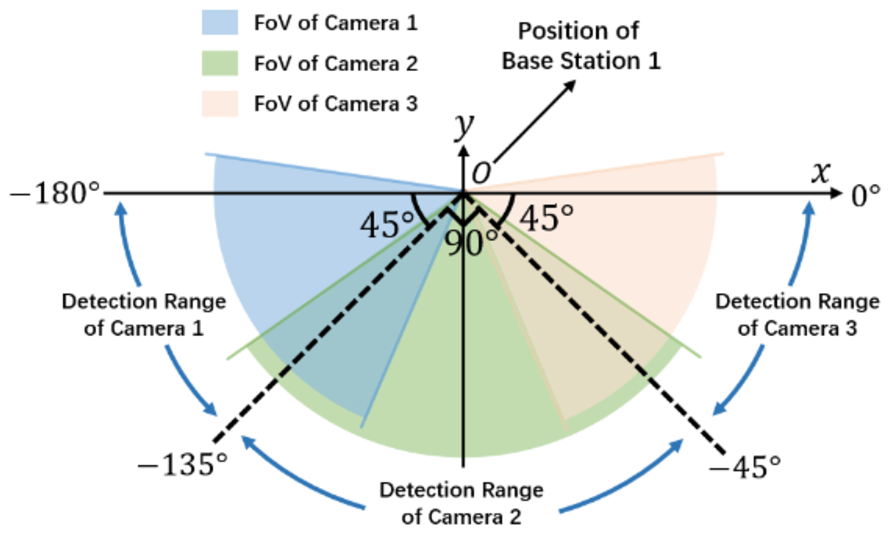}
\centering
\vspace{-0.3cm}
\caption{Detection ranges of different cameras.}
\label{fig6}
\vspace{-0.1cm}
\end{figure}


\subsubsection{Vehicle Azimuth Angle Vector Processing}
Next, we explain the use of the images collected by the BS at each time slot to generate the labels for our pretraining. In particular, we use the vehicle directional information instead of distances between the vehicles and the BS to pretrain our model, since it is hard to calculate the vehicle distances using images captured at a single location. Specifically, We first use the YOLOv4 model to detect vehicles from images in $\mathcal{I}_t$ and transform their pixel coordinates into polar coordinates according to Lemma 1. We assume that the detection ranges of different cameras are not overlapped as shown in Fig. \ref{fig6} such that a vehicle will not be captured by serveral cameras. Despite each $\boldsymbol{H}_t$ consists of the CSI matrices of $V_t$ vehicles and $\boldsymbol{q}_t$ contains the azimuth angles of $V_t^{'}$ vehicles, we do not know their corresponding relationship, which can be termed as a multi-instance multi-label learning problem according to \cite{26} and \cite{27}. To solve this problem, we make $\boldsymbol{q}_t$ into a probability distribution vector given by
\begin{equation}\label{eq:L_t}
\boldsymbol{L}_t = \left[l_{t,1},\cdots,l_{t,k},\cdots,l_{t,K}\right]\in\mathbb{R}^{1 \times K},
\end{equation}
where $l_{t,k}$ represents the probability of a vehicle at direction $k$. Specifically, we assume that the direction-finding (DF) range of our ULA is from $0^{\circ}$ to $\Omega^{\circ}$, and it is divided into $K$ intervals with a same size of $\omega=\frac{\Omega}{K}$.
Then, the probability of the azimuth angle $\phi_{t,i}$ at each direction is represented by a Gaussian distribution with $\phi_{t,i}$ and $\omega$ respectively being the mean value and standard deviation, which can be expressed as
\begin{equation}\label{eq:Weight}\begin{aligned}
\boldsymbol{w}_{t,i}=\bigg[w_{t,i,1},&\cdots,w_{t,i,k},\cdots,w_{t,i,K}\bigg]=\\
\frac{1}{\sum_{k=1}^{K} e^{-{{\frac{\left(c_k-\phi_{t,i}\right)^2}{2\omega^2}}}}}&\left[e^{-{{\frac{\left(c_1-\phi_{t,i}\right)^2}{2\omega^2}}}},\cdots,e^{-{{\frac{\left(c_K-\phi_{t,i}\right)^2}{2\omega^2}}}}\right]\in\mathbb{R}^{1 \times K},
\end{aligned}\end{equation}
where $w_{t,i,k}$ is the probability of $\phi_{t,i}$ in direction $k$, and $c_k=(k-\frac{1}{2})\omega$ denotes the center of interval $k$.
Since the distributions of different vehicles in each interval are independent, they can be considered to follow an OR logical relationship. Hence, to capture the relationship between $\boldsymbol{L}_t$ and each $\boldsymbol{w}_{t,i}$, we use the element-wise noisy-or (NOR) model here, which is given by
\begin{equation}\label{eq:Angle Interval Center}
\boldsymbol{L}_t=\mathbf{1} - \left(\mathbf{1}-\boldsymbol{w}_{t,1}\right)\odot\cdots\odot\left(\mathbf{1}-\boldsymbol{w}_{t,V_t^{'}}\right),
\end{equation}
where $\mathbf{1}=\left\{1\right\}^{1 \times K}$ is the $\mathbf{1}$-vector and $\odot$ denotes the element-wise product. 

\subsection{Components of The Proposed Positioning Scheme}
Next, we introduce the components of our designed positioning scheme which consists of three components: 1) encoder, 2) FN network $\textrm{\uppercase\expandafter{\romannumeral1}}$, 3) FN network $\textrm{\uppercase\expandafter{\romannumeral2}}$. Those components are specified as follows:

\begin{figure}[t]
\flushleft
\includegraphics[width=6cm]{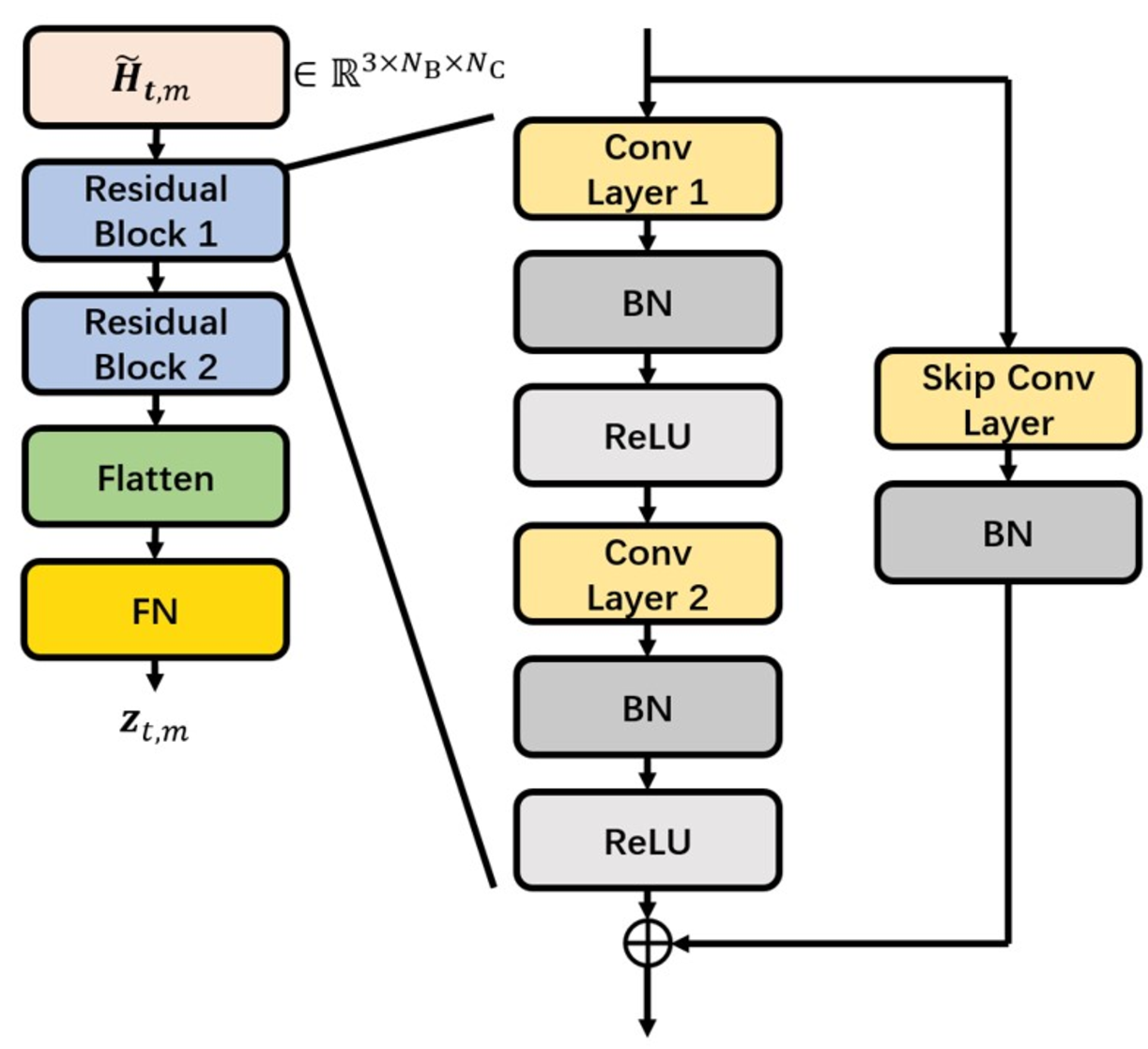}
\centering
\caption{Structure of the encoder.}
\label{fig7}
\end{figure}

\subsubsection{Encoder}
The encoder network with parameter $\boldsymbol{\xi}_{\textrm{E}}$ is used to transform each $\widetilde{\boldsymbol{H}}_{t,m}$ or $\widetilde{\boldsymbol{H}}_{j}$ into a low-dimensional feature vector \cite{28,29}. Specifically, we respectively use $\boldsymbol{z}_{t,m}\in\mathbb{R}^{1 \times N}$ or $\boldsymbol{z}_{j}\in\mathbb{R}^{1 \times N}$ to represent the feature vector of each $\widetilde{\boldsymbol{H}}_{t,m}$ or $\widetilde{\boldsymbol{H}}_{j}$ with $N$ being the output dimension of the encoder. As shown in Fig. \ref{fig7}, the encoder consists of two residual blocks (RBs) and a fully connected (FN) layer \cite{30}. Each RB consists of two convolutional layers and a skip shortcut. The output of the second RB are flattened and fed into an FN layer which will calculate the corresponding feature vector of the input CSI matrix.
\subsubsection{FN Network $\textrm{\uppercase\expandafter{\romannumeral1}}$}
\begin{figure}[t]
\centering
\includegraphics[width=6cm]{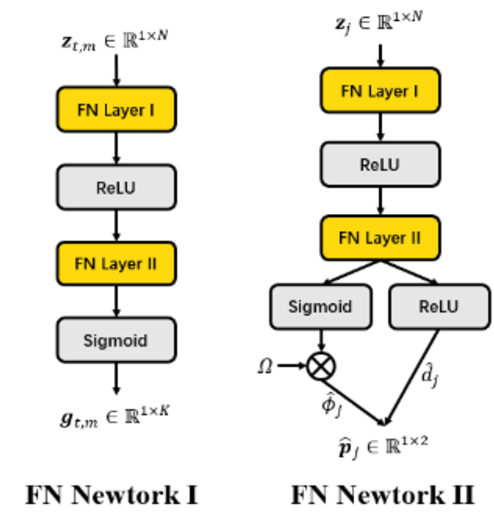}
\vspace{-0.3cm}
\caption{Structure of FN Networks.}
\label{fig8}
\end{figure}

The FN network $\textrm{\uppercase\expandafter{\romannumeral1}}$ with parameter $\boldsymbol{\xi}_{\textrm{F}_1}$ is only used in the pretraining stage. As shown in Fig. \ref{fig8}, the input of FN network $\textrm{\uppercase\expandafter{\romannumeral1}}$ is a feature vector $\boldsymbol{z}_{t,m}$ encoded from CSI matrix $\boldsymbol{H}_{t,m}$ in the pretraining dataset. The output of FN network $\textrm{\uppercase\expandafter{\romannumeral1}}$ is a predicted probability distribution in the angular domain given by
\begin{equation}\label{eq:Angle Interval Center}
\boldsymbol{g}_{t,m}=\left[g_{t,m,1},\cdots,g_{t,m,k},\cdots,g_{t,i,K}\right]\in\mathbb{R}^{1 \times K},
\end{equation}
where $g_{t,m,k}$ represents the predicted probability of the vehicle corresponding to $\boldsymbol{H}_{t,m}$ at direction $k$.

\subsubsection{FN Network $\textrm{\uppercase\expandafter{\romannumeral2}}$}
The FN network $\textrm{\uppercase\expandafter{\romannumeral2}}$ with parameter $\boldsymbol{\xi}_{\textrm{F}_2}$ is only used in the downstream training stage to predict the polar coordinate $\hat{\boldsymbol{p}}_j=\left[\hat{\phi}_j,\hat{d}_j\right]\in\mathbb{R}^{1 \times 2}$ of each $\widetilde{\boldsymbol{H}}_j$ from downstream training dataset according to its feature vector $\boldsymbol{z}_j$, where $\hat{\phi}_j\in\left[0,\Omega\right]$ is the predicted azimuth angle and $\hat{d}_j\geq 0$ is the predicted horizontal distance between a vehicle and the BS. The structure of FN Network $\textrm{\uppercase\expandafter{\romannumeral2}}$ is shown in Fig. \ref{fig8} where the activation functions of the two outputs are different.

\subsection{Training Process}

Next, we introduce the pretraining and downstream training stages in detail.
\subsubsection{Pretraining Stage}
During the pretraining, we utilize dataset $\mathcal{D}_{\textrm{P}}=\left\{ \boldsymbol{H}_{t},\boldsymbol{q}_{t}\right\}_{t=t_1}^{t_{N_{\textrm{P}}}}$ to pretrain our encoder network and the FN network $\textrm{\uppercase\expandafter{\romannumeral1}}$. Given a training data sample $\left\{ \boldsymbol{H}_{t},\boldsymbol{q}_{t}\right\}$, we first calculate $\boldsymbol{g}_{t,1},\boldsymbol{g}_{t,2},\cdots,\boldsymbol{g}_{t,V_t}$ according to $\boldsymbol{H}_{t,1},\boldsymbol{H}_{t,2},\cdots,\boldsymbol{H}_{t,V_t}$ as described in (17), and transform $\boldsymbol{q}_{t}$ into $\boldsymbol{L}_t$ by (15) and (16). Here, we consider the probability distributions of different vehicles in each horizontal direction as independent events. Hence, we can still exploit the NOR model to combine $\boldsymbol{g}_{t,1},\boldsymbol{g}_{t,2},\cdots,\boldsymbol{g}_{t,V_t}$ into one predicted probability distribution $\hat{\boldsymbol{L}}_t$, which is given by
\begin{equation}\label{eq:L_t}
\begin{aligned}
\hat{\boldsymbol{L}}_t =  \big[&\hat{l}_{t,1}\cdots,\hat{l}_{t,k},\cdots,\hat{l}_{t,K}\big]= \\ \mathbf{1} - \big(\mathbf{1}-&\boldsymbol{g}_{t,1}\big)\odot\cdots\odot\left(\mathbf{1}-\boldsymbol{g}_{t,V_t}\right)\in\mathbb{R}^{1 \times K},
\end{aligned}
\end{equation}
where $\hat{l}_{t,k}$ denotes the predicted probability of whether a vehicle may exist at direction $k$. To speed up the convergence, we use the mini-batch gradient descent (MBGD) scheme with a batch size $B_\textrm{P}$ in the pretraining stage. The goal of pretraining in iteration $n$ is to minimize the mean square error (MSE) between the ground truth and predicted probability distribution vectors, which is given by
\begin{equation}\label{eq:MSE Loss 1}
L_{\textrm{P}}^{n} =\frac{1}{B_\textrm{P}}\sum_{t\in\mathcal{B}_{\textrm{P}}^{n}} \frac{\| \boldsymbol{L}_t - \hat{\boldsymbol{L}}_t\|_2^2}{K},
\end{equation}
where $\|\cdot\|_2$ denotes the $L2$-norm of a vector. $\mathcal{B}_{\textrm{P}}^{n}$ is a dataset that contains a batch of training data samples at iteration $n$. Given the pretraining loss of iteration $n$, the parameters of the encoder network and FN network $\textrm{\uppercase\expandafter{\romannumeral1}}$ will be updated using a gradient descent method \cite{44} as follows:
\begin{equation}\label{eq:MSE Loss 1}
\boldsymbol{\xi}_{\textrm{E}}\gets\boldsymbol{\xi}_{\textrm{E}}-\lambda_{\textrm{E}}\nabla_{\boldsymbol{\xi}_{\textrm{E}}}L_{\textrm{P}}^{n}\left(\boldsymbol{\xi}_{\textrm{E}}\right),
\end{equation}
\begin{equation}\label{eq:MSE Loss 1}
\boldsymbol{\xi}_{\textrm{F}_1}\gets\boldsymbol{\xi}_{\textrm{F}_1}-\lambda_{\textrm{F}_1}\nabla_{\boldsymbol{\xi}_{\textrm{F}_1}}L_{\textrm{P}}^{n}\left(\boldsymbol{\xi}_{\textrm{F}_1}\right),
\end{equation}
where $\lambda_{\textrm{E}},\lambda_{\textrm{F}_1}$ respectively denote the learning rate of the encoder and FN network $\textrm{\uppercase\expandafter{\romannumeral1}}$. The specific pretraining algorithm is summarized in \textbf{Algorithm~1}.

\begin{algorithm}[t]\setstretch{1.1}
\small 
\caption{Pretraining stage.}
\begin{algorithmic}[1]
\STATE \textbf{Initialize:} Encoder parameter $\boldsymbol{\xi}_{\textrm{E}}$ and FN network $\textrm{\uppercase\expandafter{\romannumeral1}}$ parameter $\boldsymbol{\xi}_{\textrm{F}_1}$ generated randomly, learning rate of Encoder $\lambda_{\textrm{E}}$ and learning rate of FN network $\textrm{\uppercase\expandafter{\romannumeral1}}$ $\lambda_{\textrm{F}_1}$, the number of iteration $T$, and batch size $B_{\textrm{P}}$.
\STATE \textbf{Input:} Pretraining dataset $\mathcal{D}_{\textrm{P}}=\left\{ \boldsymbol{H}_{t},\boldsymbol{q}_{t}\right\}_{t=t_1}^{t_{N_{\textrm{P}}}}$.
\FOR {$t = t_1 \to t_{N_{\textrm{P}}}$} 
\STATE Calculate $\widetilde{\boldsymbol{H}}_t$ with $\boldsymbol{H}_{t}$ based on (11), (12) and (13).
\STATE Calculate $\boldsymbol{L}_t$ with $\boldsymbol{q}_{t}$ based on (15) and (16).
\ENDFOR

\FOR {$n = 1 \to T$} 
\STATE Randomly select $B_\textrm{P}$ collecting time slots in $\mathcal{D}_\textrm{P}$ and generate set $\mathcal{B}_{\textrm{P}}=\left\{t_{\epsilon_{n,1}},t_{\epsilon_{n,2}},\cdots,t_{\epsilon_{n,B_{\textrm{P}}}}\right\}$.
\STATE Select $\left\{\widetilde{\boldsymbol{H}}_{t_{\epsilon_{n,1}}},\boldsymbol{L}_{t_{\epsilon_{n,1}}}\right\},\cdots,\left\{\widetilde{\boldsymbol{H}}_{t_{\epsilon_{n,B_{\textrm{P}}}}},\boldsymbol{L}_{t_{\epsilon_{n,B_{\textrm{P}}}}}\right\}$ as a batch of training data.
\STATE Calculate $\hat{\boldsymbol{L}}_{t_{\epsilon_{n,1}}},\hat{\boldsymbol{L}}_{t_{\epsilon_{n,2}}},\cdots,\hat{\boldsymbol{L}}_{t_{\epsilon_{n,B_{\textrm{P}}}}}$ based on (17) and (18).
\STATE Calculate $L_{\textrm{P}}^{n}$ based on (19).
\STATE Update $\boldsymbol{\xi}_{\textrm{E}}$ by (20).
\STATE Update $\boldsymbol{\xi}_{\textrm{F}_1}$ by (21).
\ENDFOR
\end{algorithmic}
\label{algorithm_1}
\end{algorithm}

\subsubsection{Downstream Training Stage}
In this stage, we combine the pretrained encoder network with a randomly initialized FN network $\textrm{\uppercase\expandafter{\romannumeral2}}$ and retrain the encoder with dataset $\mathcal{D}_{\textrm{T}}=\left\{ \boldsymbol{H}_{j},\boldsymbol{p}_{j}\right\}_{j=1}^{N_{\textrm{T}}}$ in a fully-supervised learning manner. Since the predicted coordinate $\hat{\boldsymbol{p}}_j=\left[\hat{\phi}_j,\hat{d}_j\right]$ is a polar coordinate, we first transform each $\hat{\boldsymbol{p}}_j$ into a rectangular coordinate $\hat{\boldsymbol{p}}_j^{'}\in\mathbb{R}^{1 \times 2}$ given by
\begin{equation}\label{eq:Angle Interval Center}
\hat{\boldsymbol{p}}_j^{'}=\left[\hat{x}_j,\hat{y}_j\right]=\left[\hat{d}_j \cos{\left(\frac{\hat{\phi}_j}{180}\pi\right)},\hat{d}_j \sin{\left(\frac{\hat{\phi}_j}{180}\pi\right)}\right],
\end{equation}
where $\hat{x}_j$ and $\hat{y}_j$ respectively represent the predicted x and y coordinates. Then, we formulate the downstream training as a coordinate regression problem \cite{38}, and the MSE between ground truth and predicted position coordinates in downstream training iteration $n$ is given by
\begin{equation}\label{eq:MSE Loss}
L_{\textrm{T}}^{n} =\frac{1}{B_\textrm{T}}\sum_{j\in\mathcal{B}_{\textrm{T}}^{n}} \frac{\| \boldsymbol{p}_j - \hat{\boldsymbol{p}}_j^{'}\|_2^2}{2},
\end{equation}
where $\mathcal{B}_{\textrm{T}}^{n}$ denotes the dataset of the selected training data samples at iteration $n$, and $B_\textrm{T}$ is the number of data samples in $\mathcal{B}_{\textrm{T}}^{n}$. In the downstream training stage, we also use gradient descent method to update the parameters of the encoder and FN network $\textrm{\uppercase\expandafter{\romannumeral2}}$, which can be given as follows:
\begin{equation}\label{eq:MSE Loss 1}
\boldsymbol{\xi}_{\textrm{E}}\gets\boldsymbol{\xi}_{\textrm{E}}-\lambda_{\textrm{E}}^{'}\nabla_{\boldsymbol{\xi}_{\textrm{E}}}L_{\textrm{P}}^{n}\left(\boldsymbol{\xi}_{\textrm{E}}\right),
\end{equation}
\begin{equation}\label{eq:MSE Loss 1}
\boldsymbol{\xi}_{\textrm{F}_2}\gets\boldsymbol{\xi}_{\textrm{F}_2}-\lambda_{\textrm{F}_2}\nabla_{\boldsymbol{\xi}_{\textrm{F}_2}}L_{\textrm{P}}^{n}\left(\boldsymbol{\xi}_{\textrm{F}_2}\right),
\end{equation}
where $\lambda_{\textrm{E}}^{'}$ is the learning rate of encoder, and $\lambda_{\textrm{F}_2}$ is the learning rate of FN network $\textrm{\uppercase\expandafter{\romannumeral2}}$. Since the encoder network has been pretrained, the value of $\lambda_{\textrm{E}}^{'}$ will be much smaller than the value of $\lambda_{\textrm{E}}$. 

\subsection{Complexity of the Proposed SSL Framework}
In this section, we analyze the complexity of the proposed SSL framework. The complexity of the proposed framework is analyzed from three parts: 1) the complexity of pretraining, 2) the complexity of downstream training stage, and 3) the complexity of inference. Specifically, the pretraining process and downstream training process will be respectively conducted once in each iteration during the offline phase, while the inference process of predicting vehicle positions is conducted for each vehicle during the online phase. Next, we first introduce the complexity of pretraining. Then, we explain the complexity of downstream training. Finally, we introduce the complexity of inference.
\subsubsection{Complexity of Pretraining}
The complexity of pretraining lies in computing the feature vector $\boldsymbol{z}_{t,m}$ using the encoder and calculating the distribution vector $\hat{\boldsymbol{L}}_t$ using FN network $\textrm{\uppercase\expandafter{\romannumeral1}}$. Since the encoder network consists of two RBs and a FN layer, the complexity of feature vector computation depends on the input and output channels of the convolutional layers in each RB, the spatial sizes of the convolutional kernels, the output widths and lengths of each convolutional layer, and the output dimension of the FN layer. According to \cite{32}, the complexity of feature vector computation is given by
\begin{equation}\label{eq:MSE Loss 1}
\centering
\begin{aligned}
\mathcal{O}\Bigg(B_{\textrm{P}}& \overline{V}\Bigg(\sum_{i=1}^{2}\bigg( \sum_{j=1}^{2}  \big( \big( c_{i,j-1}  s_{i,j}^2 +2 \big) c_{i,j}  \alpha_{i,j}\beta_{i,j} \big) \\ + &\big(c_{i,0} {s_i^{*}}^2  +2\big) c_{i,2}\alpha_{i,2}\beta_{i,2} \bigg)+\alpha_{2,2}\beta_{2,2}N \Bigg)\Bigg)   \\ 
=&\mathcal{O}\Bigg(\mathop{\max}\limits_{i,j} \, \bigg(\left( c_{i,j-1}  s_{i,j}^2 +2 \right) c_{i,j} \bigg)\Bigg),
\end{aligned}
\end{equation}
where $c_{i,j}$ and $s_{i,j}$ respectively denote the output channel and kernel size of convolutional layer $j$ in RB $i$, $c_{i,0}$ is the input channel of the first convolutional layer in RB $i$, $\alpha_{i,j},\beta_{i,j}$ are the output width and length of convolutional layer $j$ in RB $i$, ${s_i^{*}}$ is the skip convolutional layer kernel size of RB $i$, and $\overline{V}=\frac{1}{N_{\textrm{P}}}\sum_{t=t_1}^{t_{N_{\textrm{P}}}} V_t$ denote the average number of CSI matrices in each $\boldsymbol{H}_t$. Given the feature vectors, the complexity of calculating the vehicle distribution vectors in each horizontal direction using FN network $\textrm{\uppercase\expandafter{\romannumeral1}}$ is $\mathcal{O}\left(B_{\textrm{P}} \overline{V}\left(N M_{\textrm{F}_1} + K M_{\textrm{F}_1}\right)\right)=\mathcal{O}\left(M_{\textrm{F}_1}\left(N  + K \right)\right)$, where $M_{\textrm{F}_1}$ represents the number of neurons in the first FN layer of FN network $\textrm{\uppercase\expandafter{\romannumeral1}}$. Therefore, the computational complexity of the pretraining stage is given by
\begin{equation}\label{eq:MSE Loss 1}
\begin{split}
\mathcal{O}\Bigg(\mathop{\max}\limits_{i,j} \, \bigg(\left( c_{i,j-1}  s_{i,j}^2 +2 \right) c_{i,j} \bigg)+M_{\textrm{F}_1}\big(N  + K\big)\Bigg).
\end{split}
\end{equation}

\begin{figure}[t]
\centering
\includegraphics[width=8.5cm]{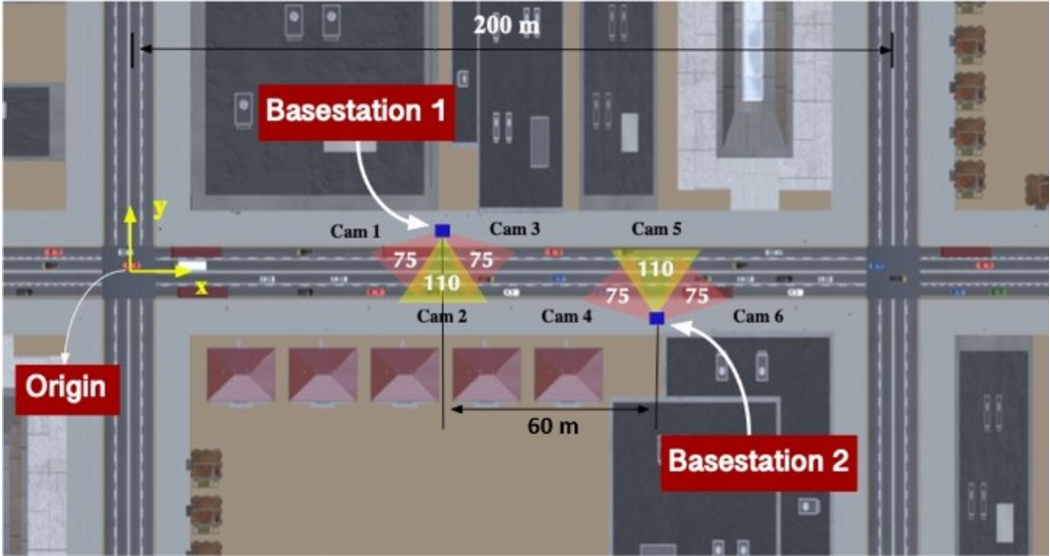}
\centering
\vspace{-0.1cm}
\caption{Scenario of the dataset \cite{33}.}
\label{fig9}
\end{figure}

\subsubsection{Complexity of Downstream Training}
The complexity of the downstream training stage lies in computing the feature vector $\boldsymbol{z}_{j}$ using encoder and determining the polar coordinate $\hat{\boldsymbol{p}}_j$ with FN network $\textrm{\uppercase\expandafter{\romannumeral2}}$. Therefore, the computational complexity is
\begin{equation}\label{eq:MSE Loss 1}
\begin{split}
\mathcal{O}\Bigg(\mathop{\max}\limits_{i,j} \, \bigg(\left( c_{i,j-1}  s_{i,j}^2 +2 \right) c_{i,j} \bigg)+M_{\textrm{F}_2}\big(N  + K\big)\Bigg),
\end{split}
\end{equation}
where $M_{\textrm{F}_2}$ represents the number of neurons in the first FN layer of FN network $\textrm{\uppercase\expandafter{\romannumeral2}}$.

\subsubsection{Complexity for Inference}
The complexity of inference lies in predicting the vehicle polar coordinates using encoder and FN network $\textrm{\uppercase\expandafter{\romannumeral2}}$. Compared with downstream training, we do not need to calculate the variance and mean value of each batch of data. Hence, the inference complexity of predicting the position of a vehicle is 
\begin{equation}\label{eq:MSE Loss 1}
\begin{split}
\mathcal{O}\Bigg(\mathop{\max}\limits_{i,j} \, \bigg(\left( c_{i,j-1}  s_{i,j}^2 +1 \right) c_{i,j} \bigg)+M_{\textrm{F}_2}\big(N  + K\big)\Bigg).
\end{split}
\end{equation}

    

\section{Simulation results and analysis}
Next, we evaluate the performance of our proposed scheme on a public dataset called vision-wireless (ViWi) dataset \cite{33}. We first introduce the ViWi dataset. Then, we describe the baselines used in our experiments. Finally, we analyze the performance of our proposed scheme.

\begin{table}\footnotesize
\newcommand{\tabincell}[2]{\begin{tabular}{@{}#1@{}}#1.1\end{tabular}}
\renewcommand\arraystretch{1.15}
\caption[table]{{System Parameters \& Simulation Settings}}
\centering
\begin{tabular}{|c|c|c|c|c|c|}
\hline
\!\textbf{Parameter}\! \!\!& \textbf{value} &\! \textbf{Parameter} \!& \textbf{Value} \\
\hline
$C$ & 3 &  $\Omega$ & $180^{\circ}$\\
\hline
$W_c \left(c=1,2,3\right)$ & 1280 &  $N$ & 32\\
\hline
$H_c \left(c=1,2,3\right)$ & 720 & $K$ & 15,30,60 \\
\hline
$N_\textrm{B}$ & 16 & $B_\textrm{P}$ & 64 \\
\hline
$N_\textrm{C}$ & 52 & $B_\textrm{T}$ & 32 \\
\hline
$N_{\textrm{P}}$ & 3000 & $\lambda_{\textrm{E}},\lambda_{\textrm{F1}},\lambda_{\textrm{F2}}$ & $10^{-3}$ \\
\hline  
$N_{\textrm{V}}$ & 1480 & $\lambda_{\textrm{E}}^{'}$ & $\frac{1}{20}\times 10^{-3}$ \\
\hline
$N_{\textrm{T}}$ & \makecell{200,300,500, \\ 750,1000} & &  \\
\hline
\end{tabular}
\label{table2}
\end{table}

\subsection{Dataset and baselines}
As shown in Fig. \ref{fig9}, the datasets used in simulations are collected from a downtown scenario with multiple served vehicles and two BSs located at each side of the street. All vehicles are located within a 180-degree range in front of each BS. Each BS is equipped with three differently-oriented cameras such that their FoVs can cover the whole street. We only use the dataset collected from the vehicles that are captured by BS 1. The distances between BS 1 and these vehicles are no more than 40 m. Each sample consists of three images taken by three cameras, and the CSI matrices and position coordinates of the served vehicles. Since the time slots of two successive samples are very close, we only take one sample from every four samples. After down-sampling and filtering out the unusable samples, we finally select 4000 samples from the dataset. 3000 samples are randomly selected as the pretraining dataset $\mathcal{D}_{\textrm{P}}$. The remaining 1000 samples are used as the labeled CSI dataset $\mathcal{D}_{\textrm{T}}$ and validation dataset. During pretraining, the number of iteration is 3000 and the initial learning rates will be reduced to their 90\% by every 100 iterations. In downstream training stage, the model is retrained by 3000 epochs and the learning rates will be reduced to 90\% by every 100 epochs. Here, one epoch includes $\frac{N_{\textrm{P}}}{B_\textrm{P}}$ iterations since an epoch represents the process that the model is once trained by the whole dataset, while an iteration is the process that the model parameters are updated by a mini-batch of data samples. The frequency and bandwidth used for vehicle positioning are 28 GHz and 0.2 GHz. Other system parameters and simulation settings are listed in Table \ref{table2}. 

For comparison purposes, we consider two baselines. In baseline a), the encoder is randomly initialized and trained only by the labeled dataset $\mathcal{D}_{\textrm{T}}$ without using the unlabeled dataset $\mathcal{D}_{\textrm{P}}$. Since the encoder of baseline a) is not pretrained, both the initial learning rates of the encoder and FN network $\textrm{\uppercase\expandafter{\romannumeral2}}$ are $10^{-3}$. The number of epochs to train the baseline model is 3000 and the learning rate will be reduced to its 90\% by every 100 epochs. In baseline b), the encoder is pretrained with the self-supervised contrastive learning (SSCL) scheme \cite{25} using the unlabeled dataset. SSCL is a an unsupervised learning paradigm which is commonly used for model pretraining in semi-supervised learning algorithms. Specifically, in SSCL, the models are pretrained by contrasting positive and negative samples derived from the unlabeled dataset. Here, one positive sample is an unlabeled CSI with Gaussian noise, and its negative samples are other unlabeled CSI samples with noise in the dataset. The reason that we compare the proposed scheme with baseline b) is to show the superiority of using image to pretrain the model. We do not compare the proposed method with algorithms that only use image data to localize vehicles since our designed method does not need to use images for vehicle positioning during the implementation stage.

\subsection{Evaluation Metrics}
In our experiment, we measure the performances of pretraining and downstream training by calculating their evaluation metrics on a validation dataset $\mathcal{D}_{\textrm{V}}=\left\{ \boldsymbol{H}_{j},\boldsymbol{p}_{j}\right\}_{j=1}^{N_{\textrm{V}}}$ that includes $N_{\textrm{V}}$ labeled samples.

In pretraining stage, we use the azimuth angle error between predicted and ground truth directions to measure the performance of our model. Given CSI sample $j$ from $\mathcal{D}_{\textrm{V}}$, the azimuth angle error is defined as 
\begin{equation}\label{eq:Metric}
s_{j} =\omega \cdot \left| r_{j}-\hat{r}_{j} \right|,
\end{equation}
where $r_{j} =\lceil \frac{\phi_j}{\omega} \rceil$ is the ground truth direction and $\lceil \cdot \rceil$ denotes the round up operation, $\hat{r}_{j} =\underset{k=1,\cdots,K}{{\arg\max} \, g_{j,k}}$ represents the predicted direction of the vehicle.

\begin{figure}[t]
\centering
\includegraphics[width=8.5cm]{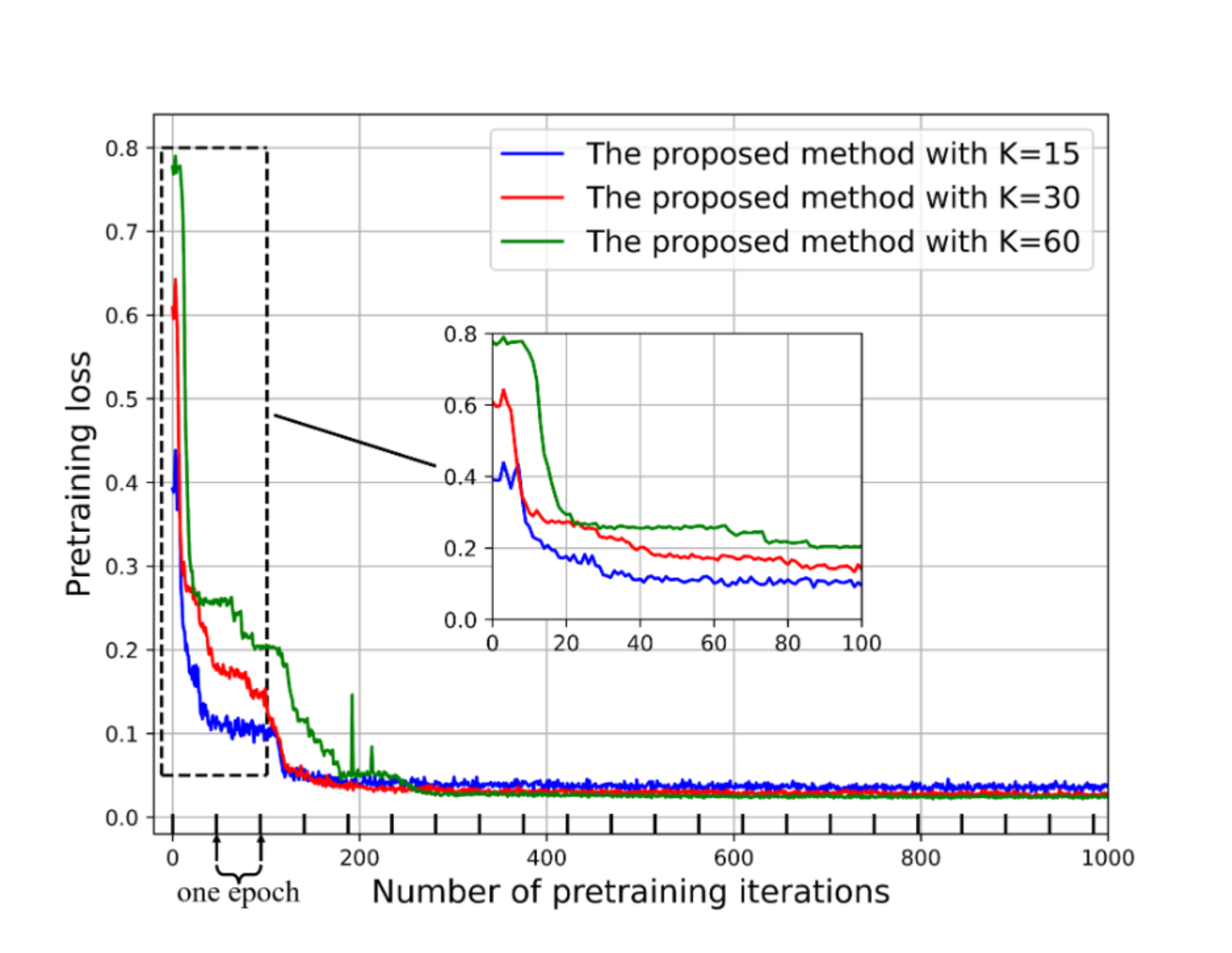}
\centering
\vspace{-0.1cm}
\caption{The convergence of proposed algorithm with different $K$.}
\label{fig10}
\end{figure}

\begin{figure}[t]
\centering
\subfigbottomskip=2pt
\subfigcapskip=-10pt
\subfigure[CDF of azimuth angle error with different $K$.]{\includegraphics[width=8.5cm]{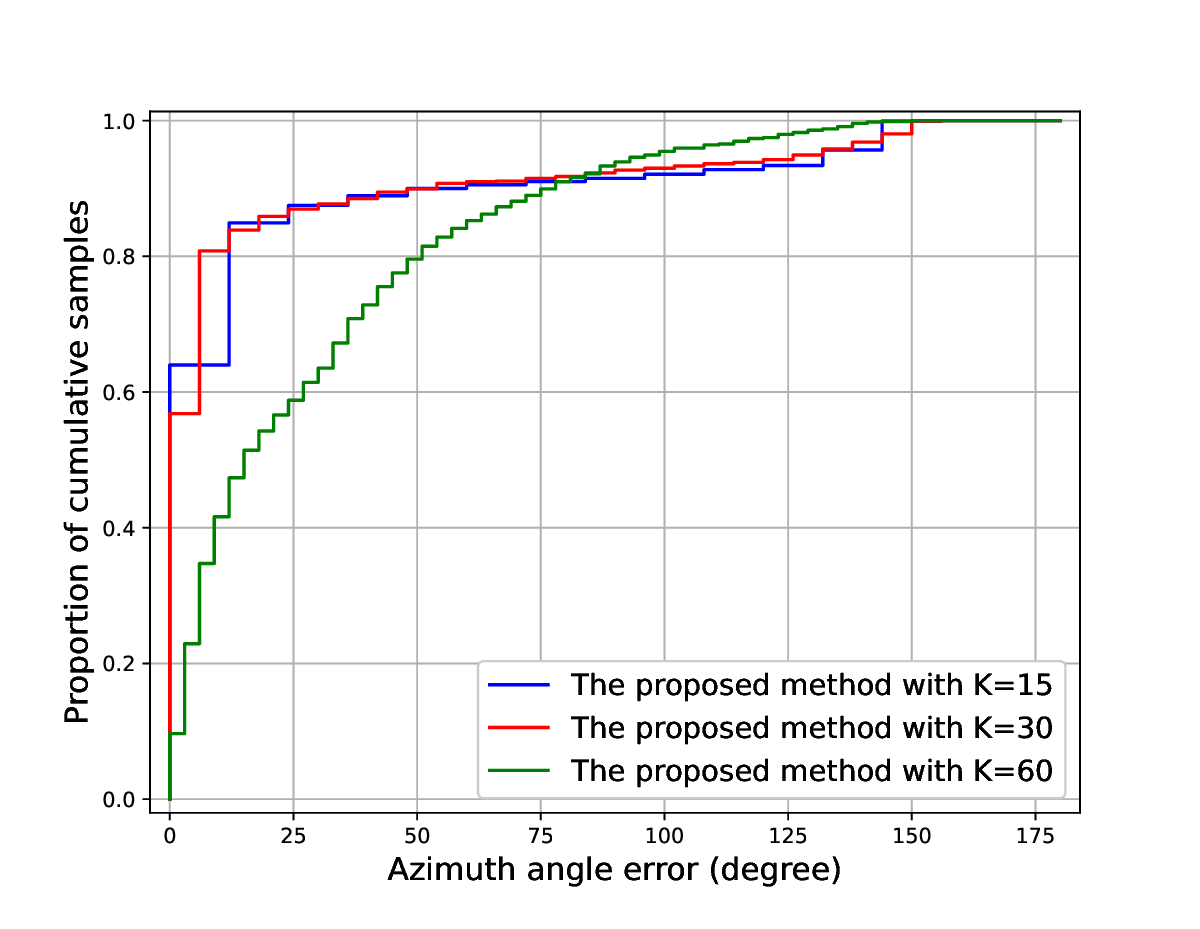}}
\subfigure[CDF of azimuth angle error with different amount of pretraining data ($K=30$).]{\includegraphics[width=8.5cm]{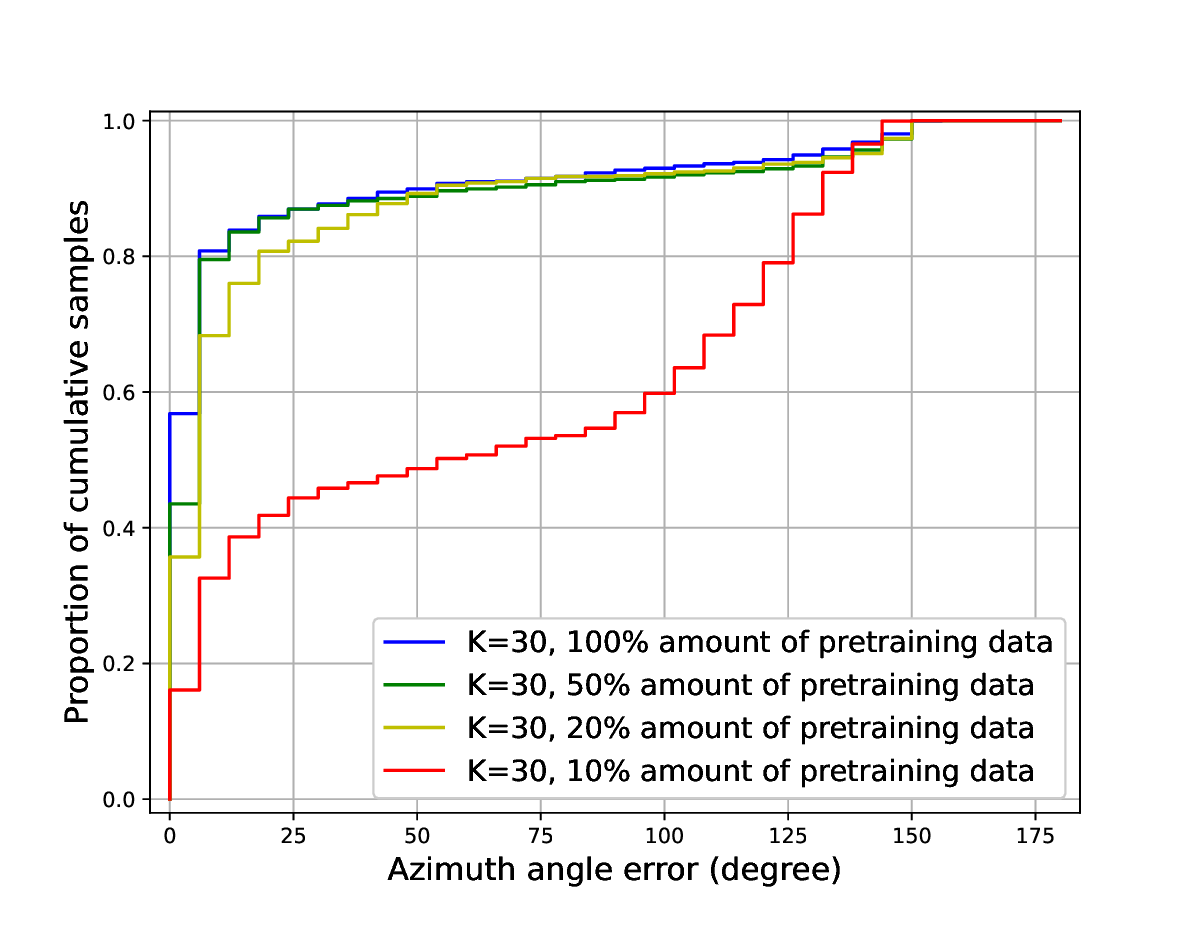}}\vspace{0.1cm}
\caption{CDF of predicted and ground truth azimuth angle error with different pretraining settings.}
\label{fig11}
\end{figure}

\begin{figure*}[t]
\centering
\subfigbottomskip=2pt
\subfigcapskip=-10pt
\subfigure[Example with $K=15$.]{\includegraphics[width=8.5cm]{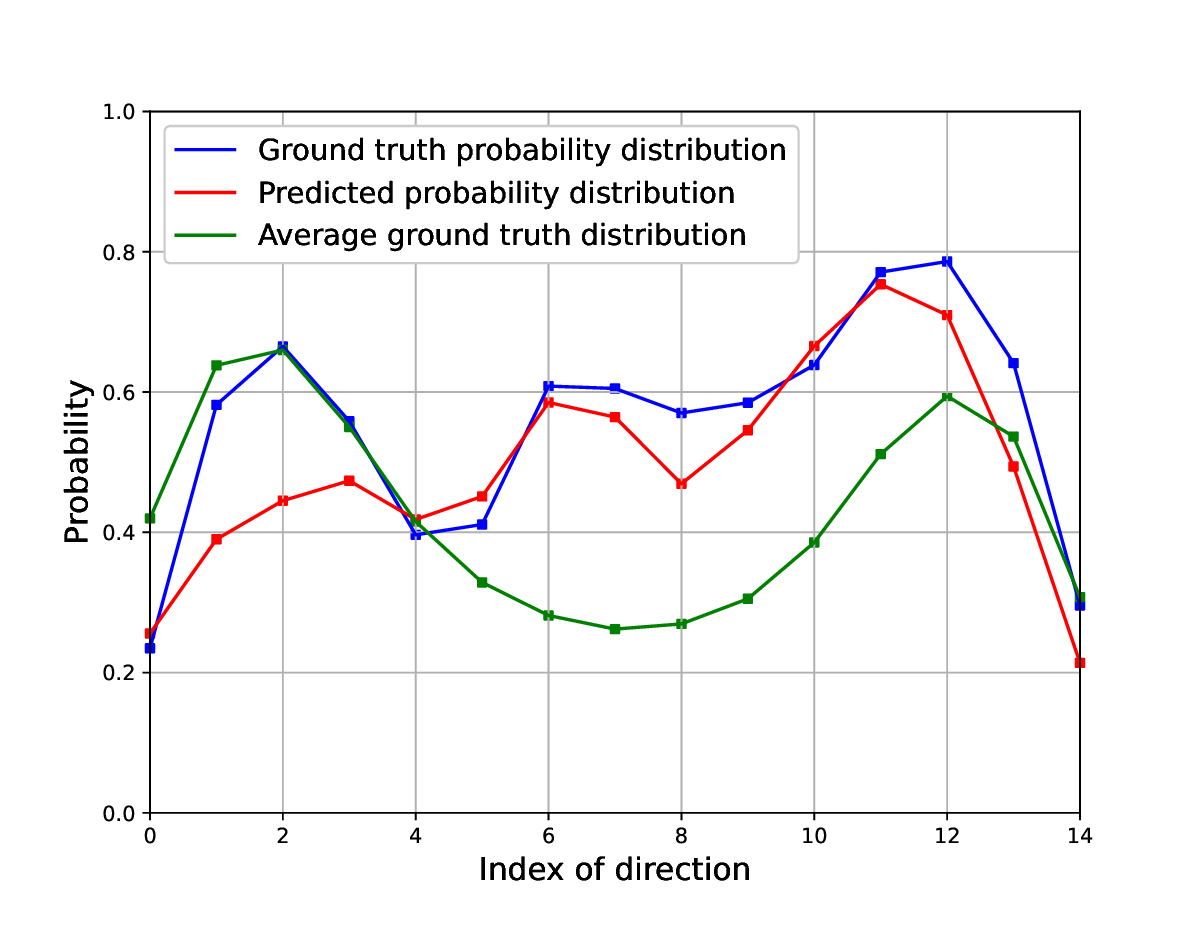}}
\subfigure[Example with $K=30$.]{\includegraphics[width=8.5cm]{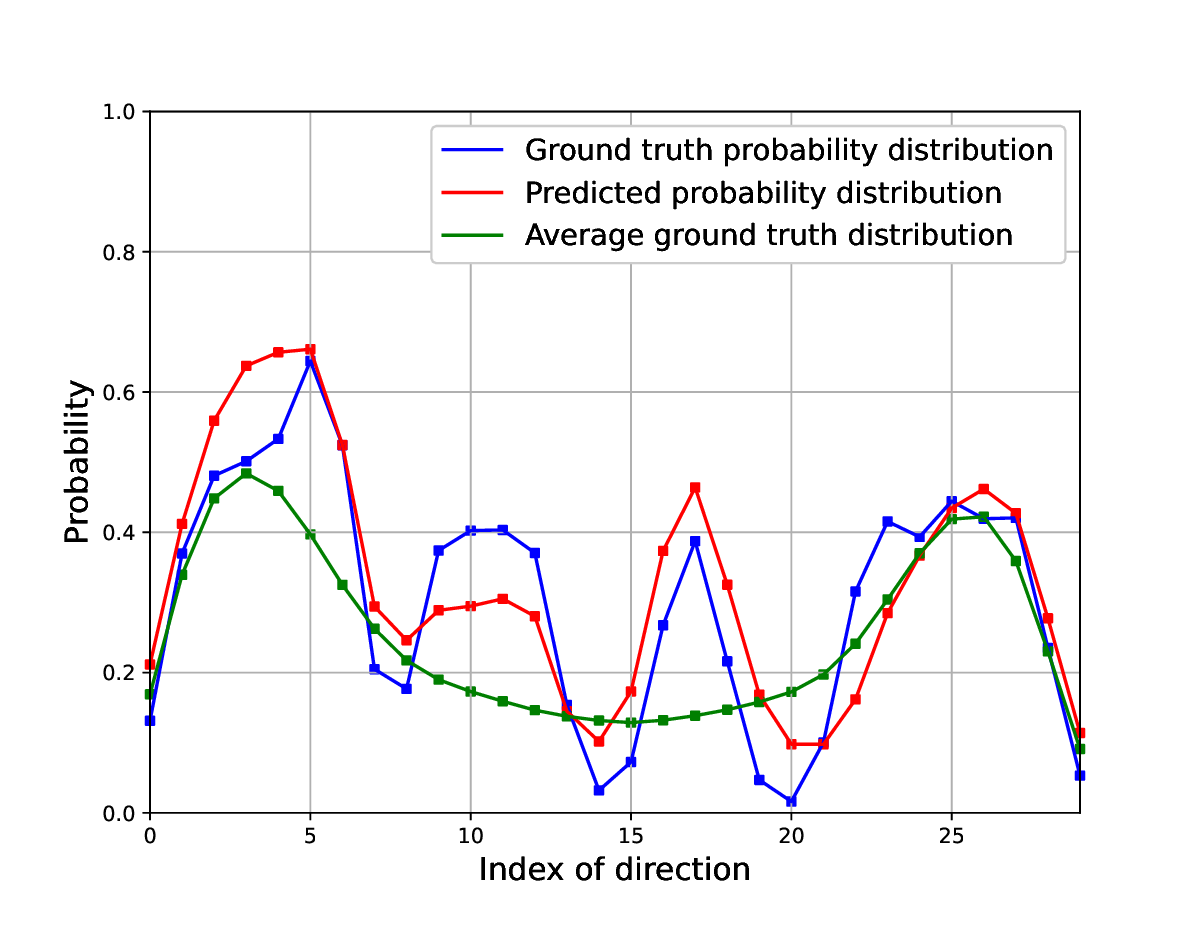}}\vspace{-0.3cm}
\quad
\subfigure[Example 1 with $K=60$.]{\includegraphics[width=8.5cm]{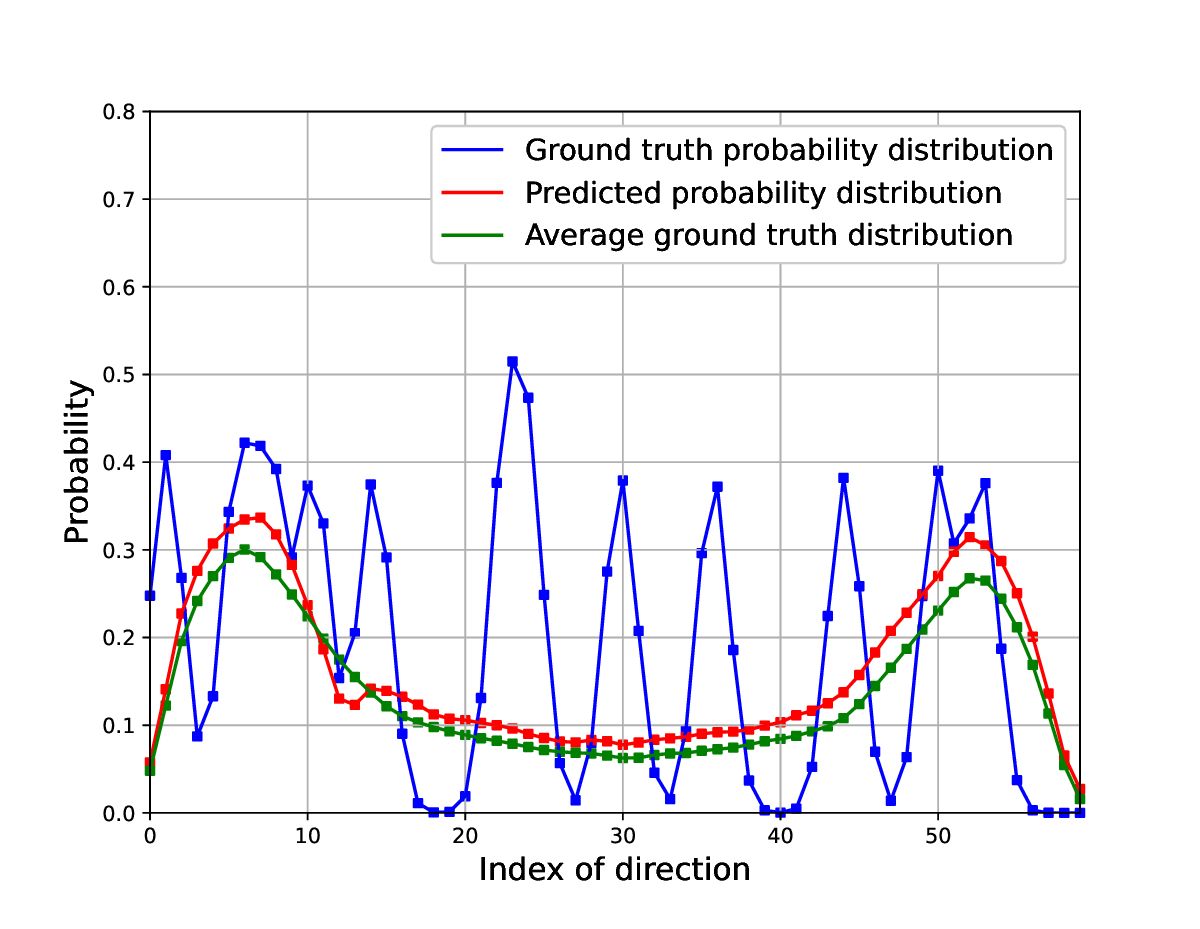}}
\subfigure[Example 2 with $K=60$.]{\includegraphics[width=8.5cm]{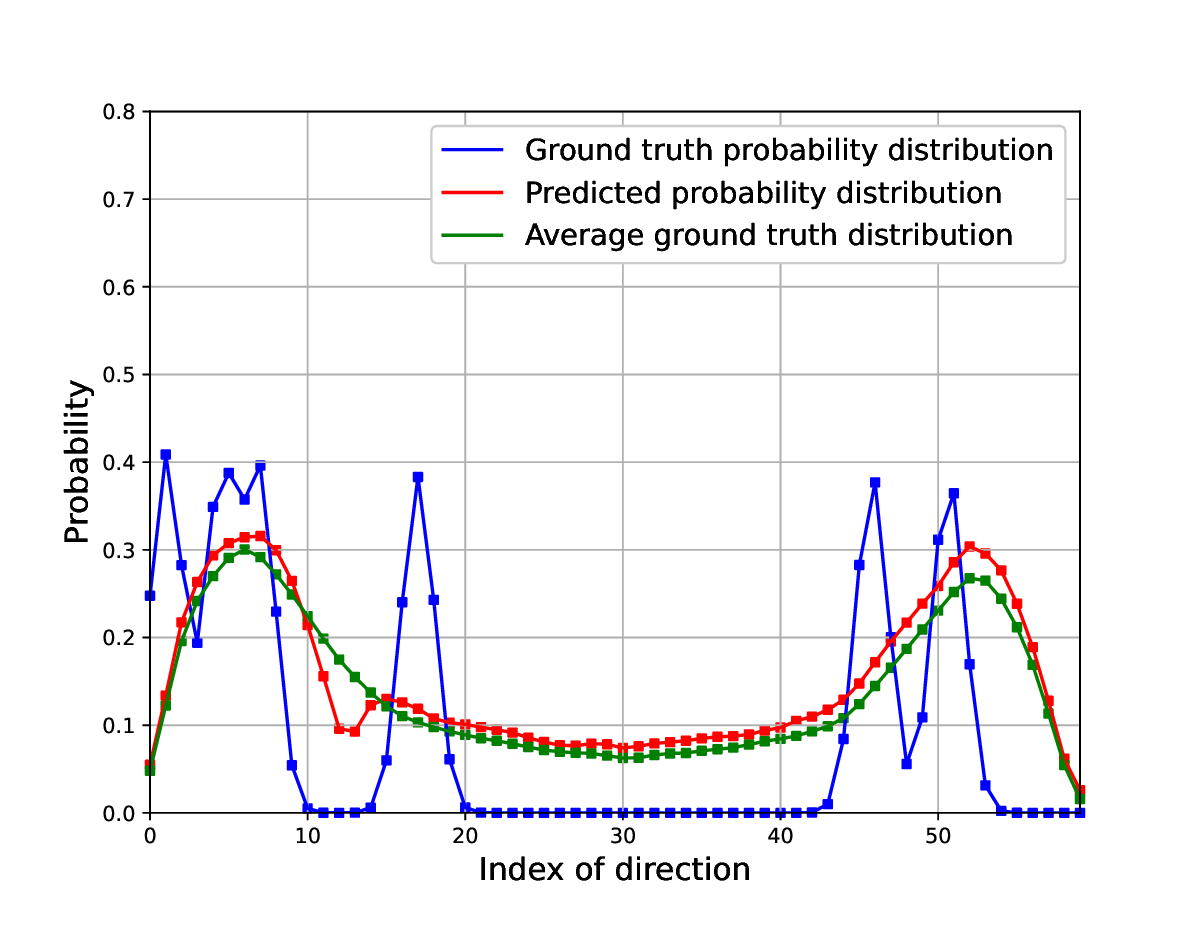}}
\caption{Examples of ground truth and predicted probability distributions with different $K$.}
\label{fig12}
\vspace{-0.1cm}
\end{figure*}

In the downstream training stage, we first use mean absolute error (MAE) \cite{34} to measure the performances of vehicle azimuth angle predictions and the predictions of distances between vehicles and the BS of the considered methods, as follows:
\begin{equation}\label{eq:Metric}
S_{\textrm{MAE}}^{\textrm{A}} =\frac{1}{N_{\textrm{V}}}\sum_{j=1}^{N_{\textrm{V}}} \left|\phi_j-\hat{\phi}_j \right|,
\end{equation}
\begin{equation}\label{eq:Metric}
S_{\textrm{MAE}}^{\textrm{D}} =\frac{1}{N_{\textrm{V}}}\sum_{j=1}^{N_{\textrm{V}}} \left|d_j-\hat{d}_j \right|.
\end{equation}
Then, to evaluate positioning accuracy, we use the mean Euclidean distance between the predicted and ground truth positions as the positioning error to evaluate the positioning accuracy \cite{37}, which is given by $S_{\textrm{pos}} =\frac{1}{N_{\textrm{V}}}\sum_{j=1}^{N_{\textrm{V}}} \| \boldsymbol{p}_j - \hat{\boldsymbol{p}}_j^{'}\|_2$. Since

\subsection{Performance Evaluation}
\subsubsection{Pretraining}
In Fig. \ref{fig10}, we show how the pretraining loss changes as the number of training iterations increases. From Fig. \ref{fig10}, we see that the pretraining loss of all considered methods remain unchanged when the number of iterations is larger than 250 which implies that all considered algorithms converge after 250 iterations. We can also observe that the pretraining loss of the proposed algorithm with $K=60$ has the largest initial loss and the lowest descending rate, while the proposed algorithm with $K=15$ has the smallest initial loss and the highest descending rate. This is because as $K$ increases, the number of components in the predicted probability distribution vector increases, which may increase the difficulties of finding the optimal model that can accurately predict the directions of vehicles.


Fig. \ref{fig11} shows cumulative distribution function (CDF) of the azimuth angle errors in validation dataset resulting from the proposed algorithms with different configurations. In Fig. \ref{fig11} (a), we show CDF of the proposed algorithms with the number of angle intervals $K=15,30,60$ and are trained by the whole pretraining dataset. From Fig. \ref{fig11} (a), we see that 47\% of azimuth angle errors resulting from the algorithm with $K=60$ are below $12^{\circ}$, and 85\% of azimuth angle errors of the algorithm with $K=15$ or $30$ are lower than $12^{\circ}$. This is because when $K=60$, the proposed algorithm might fail to find the globally optimal model, but converge to a locally optimal solution. From Fig. \ref{fig11} (a), we also see that 82\% of azimuth angle errors of the algorithm with $K=30$ are below $6^{\circ}$, but only 64\% of azimuth angle errors of the algorithm with $K=15$ are lower than $6^{\circ}$. This is because as $K$ grows from $15$ to $30$, the size of an azimuth angle interval $\omega=\frac{\Omega}{K}$ in the probability distribution vector will decrease from $12^{\circ}$ to $6^{\circ}$. Hence, the accuracy of azimuth angle prediction resulting from the proposed method increases. In Fig. \ref{fig11} (b), we show the CDF of the proposed algorithms ($K=30$) that are trained by 100\%, 50\%, 20\%, and 10\% amount of pretraining data. From Fig. \ref{fig11} (b), we observe that 31\% of azimuth angle errors of the algorithm that is trained by 10\% of pretraining data are lower than $12^{\circ}$. Meanwhile, 85\% of azimuth angle errors of the algorithm that is trained by the whole pretraining dataset are below $12^{\circ}$. This is because the model is overfitting on the pretraining dataset when the amount of pretraining data is small \cite{35}. Therefore, the accuracy of azimuth angle prediction resulting from the proposed method decreases.

In Fig. \ref{fig12}, we verify the statement that the pretraining algorithm with $K=60$ might converge to a locally optimal solution rather than the globally optimal model. In this figure, we show the ground truth distribution vector $\boldsymbol{L}_t$, the predicted distribution vector $\hat{\boldsymbol{L}}_t$, and the average ground truth distribution vector $\overline{\boldsymbol{L}}=\frac{1}{N_{\textrm{P}}}\sum_{t=t_1}^{t_{N_{\textrm{P}}}} \boldsymbol{L}_t$. If the algorithm converges to a globally optimal solution, the predicted distribution vector $\hat{\boldsymbol{L}}_t$ should be close to its ground truth distribution vector $\boldsymbol{L}_t$ since the target of the pretraining algorithms is to minimize the error between each $\hat{\boldsymbol{L}}_t$ and $\boldsymbol{L}_t$ in $\mathcal{D}_{\textrm{P}}$. Figs. \ref{fig12} (a) and (b) show that the $\hat{\boldsymbol{L}}_t$ is very close to $\boldsymbol{L}_t$ when the model is pretrained by the algorithm with $K=15$ or $30$. This is because the algorithm with $K=15$ or $30$ converges to a globally optimal model that can accurately predict the direction of each vehicle. In contrast, in Figs. \ref{fig12} (c) and (d), $\hat{\boldsymbol{L}}_t$ are both very close to the average ground truth distibution vector $\overline{\boldsymbol{L}}$ rather than $\boldsymbol{L}_t$. This is becase when $K=60$, the difficulty of finding the globally optimal solution increases. Therefore, the algorithm with $K=60$ converges to a locally optimal model which maps all the different input data to the same average ground truth distribution vector $\overline{\boldsymbol{L}}$. Since the pretraining parameters (i.e. $K$ and $N_{\textrm{P}}$) can affect the number of labeled CSI samples required by the proposed method to reach a certain performance in downstream training stage, we select the parameters that yield the best pretraining results for downstream training (i.e. $K=30$, $N_{\textrm{P}}=3000$).

\begin{figure}[t]
\centering
\includegraphics[width=8.5cm]{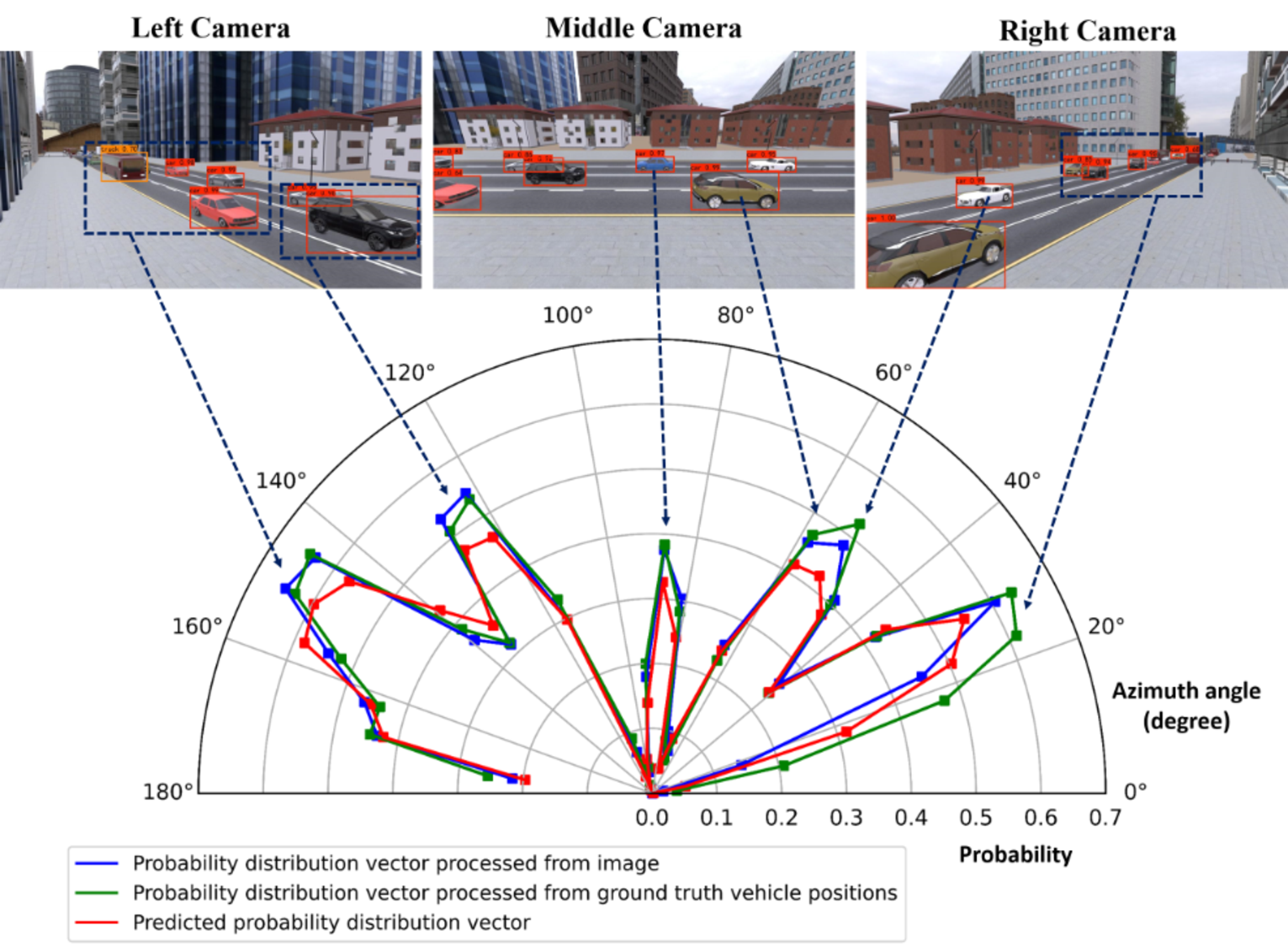}
\centering
\vspace{-0.1cm}
\caption{Probability distribution vectors and detected vehicles in corresponding images.}
\label{fig13}
\vspace{-0.1cm}
\end{figure}

In Fig. \ref{fig13}, we show an example of azimuth angle prediction in validation dataset. Specifically, we show the vehicle detection results of the images, the probability distribution vector obtained from images, the probability distribution vector calculated via ground truth positions of vehicles, and the predicted distribution vector. From Fig. \ref{fig13}, we can see that the probability distribution vector obtained from images is very close to the probability distribution vector resulting from the ground truth positions of vehicles, which verifies the feasibility of using image data to generate labels for unlabeled CSI data. We can also observe from Fig. \ref{fig13} that the predicted probability distribution vector is similar to the other two vectors. This implies that the proposed pretraining method can effectively train the model to predict vehicle azimuth angles without knowing the concrete corresponding relationship between each azimuth angle and CSI.

\subsubsection{Downstream training}
\begin{figure}[t]
\centering
\includegraphics[width=8.5cm]{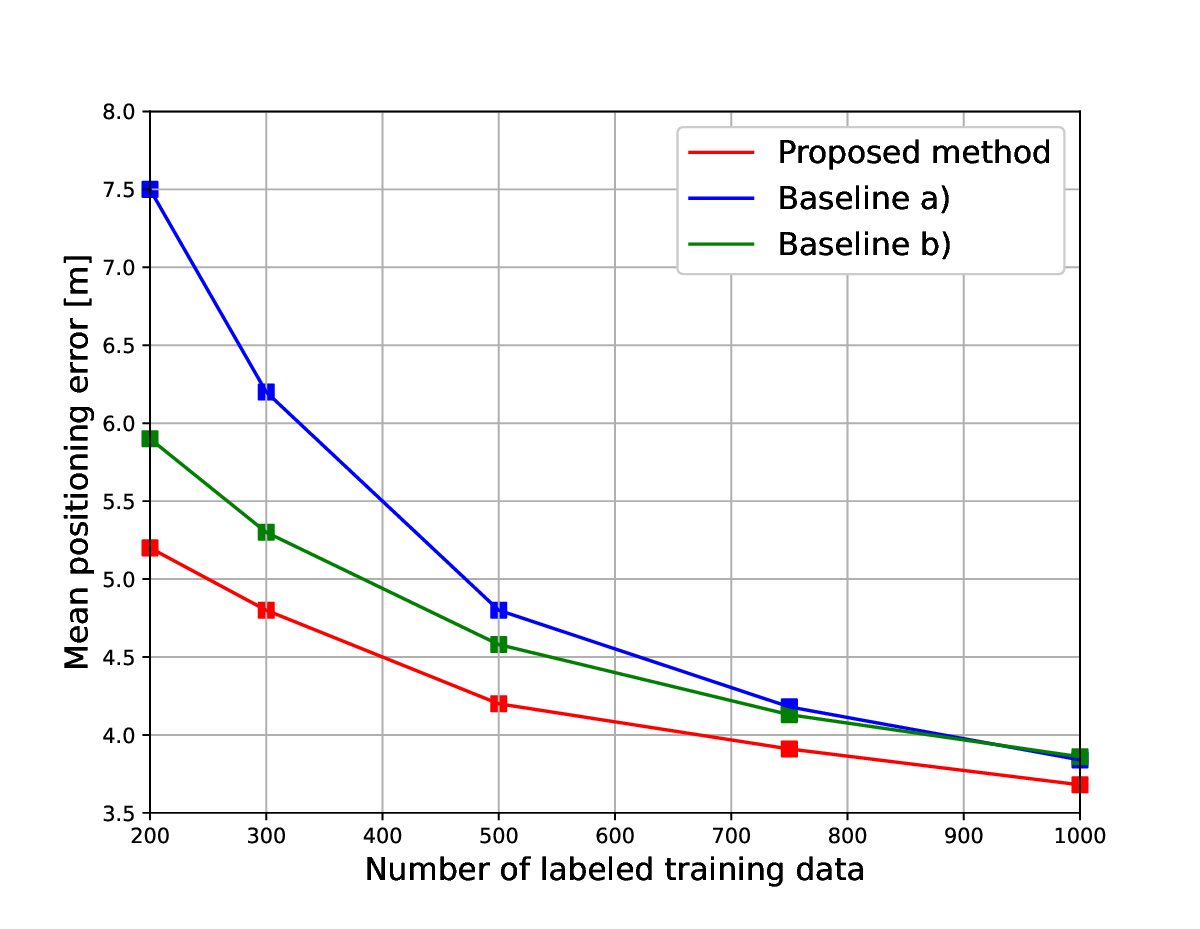}
\centering
\vspace{-0.1cm}
\caption{Positioning error with different number of data in downstream training dataset.}
\label{fig14}
\vspace{-0.1cm}
\end{figure}

In Fig. \ref{fig14}, we show how the mean positioning error changes as the number $N_{\textrm{T}}$ of labeled data samples in the downstream training dataset varies. From this figure, we can see that the proposed algorithm can effectively improve the positioning accuracy especially when the amount of labeled training data samples is very small. Specifically, under the current experimental setup of the pretraining stage, the appropriate number of labeled CSI samples to reach the optimal performance is 1000, since the positioning error of the proposed method is similar to baseline a) when the number of labeled CSI is larger than 1000. We see that the proposed algorithm can reduce the mean positioning error by up to 30\% when $N_{\textrm{T}}=200$ compared to baseline a). This is because the proposed method pretrains the positioning model with a large amount of unlabeled data such that the pretrained model has a better generalization capacity. We can also observe that the proposed method can reduce the mean positioning error by up to 9.5\% when $N_{\textrm{T}}=200$ compared to baseline b). This stems from the fact that the pretraining labels of the proposed method contain the information of vehicle directions which is strongly correlated to the locations of vehicle. In contrast, the pretraining target of baseline b) is to minimize the contrastive loss between unlabeled CSI data, which is not related to vehicle positioning. As a result, the encoder of the proposed method can better extract features related to vehicle locations, and thus achieving lower positioning error.

\begin{figure}[t]
\centering
\subfigbottomskip=2pt
\subfigcapskip=-10pt
\subfigure[Validation MAE of predicted angle with $N_{\textrm{T}}=300$.]{\includegraphics[width=8.5cm]{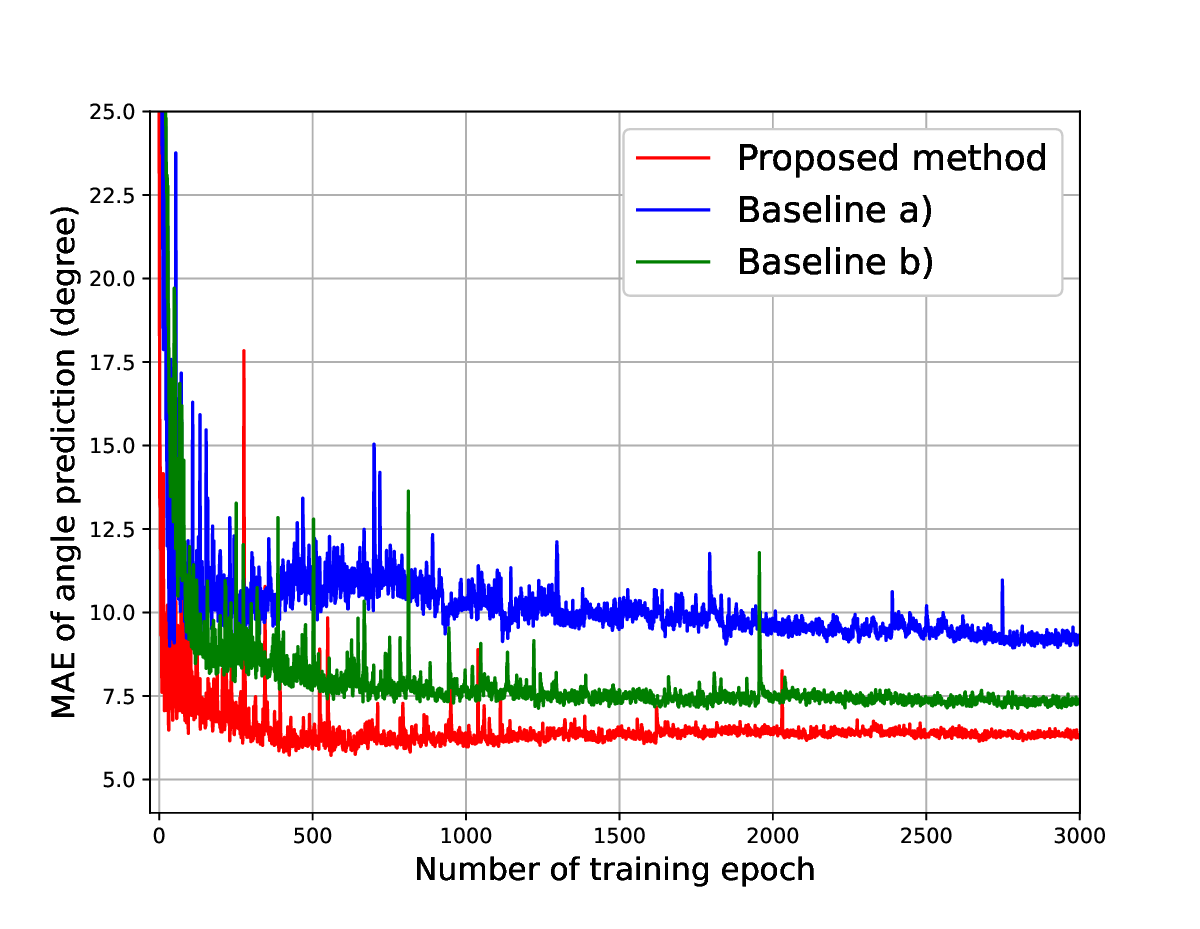}}\vspace{-0.3cm}
\subfigure[Validation MAE of predicted distance with $N_{\textrm{T}}=300$.]{\includegraphics[width=8.5cm]{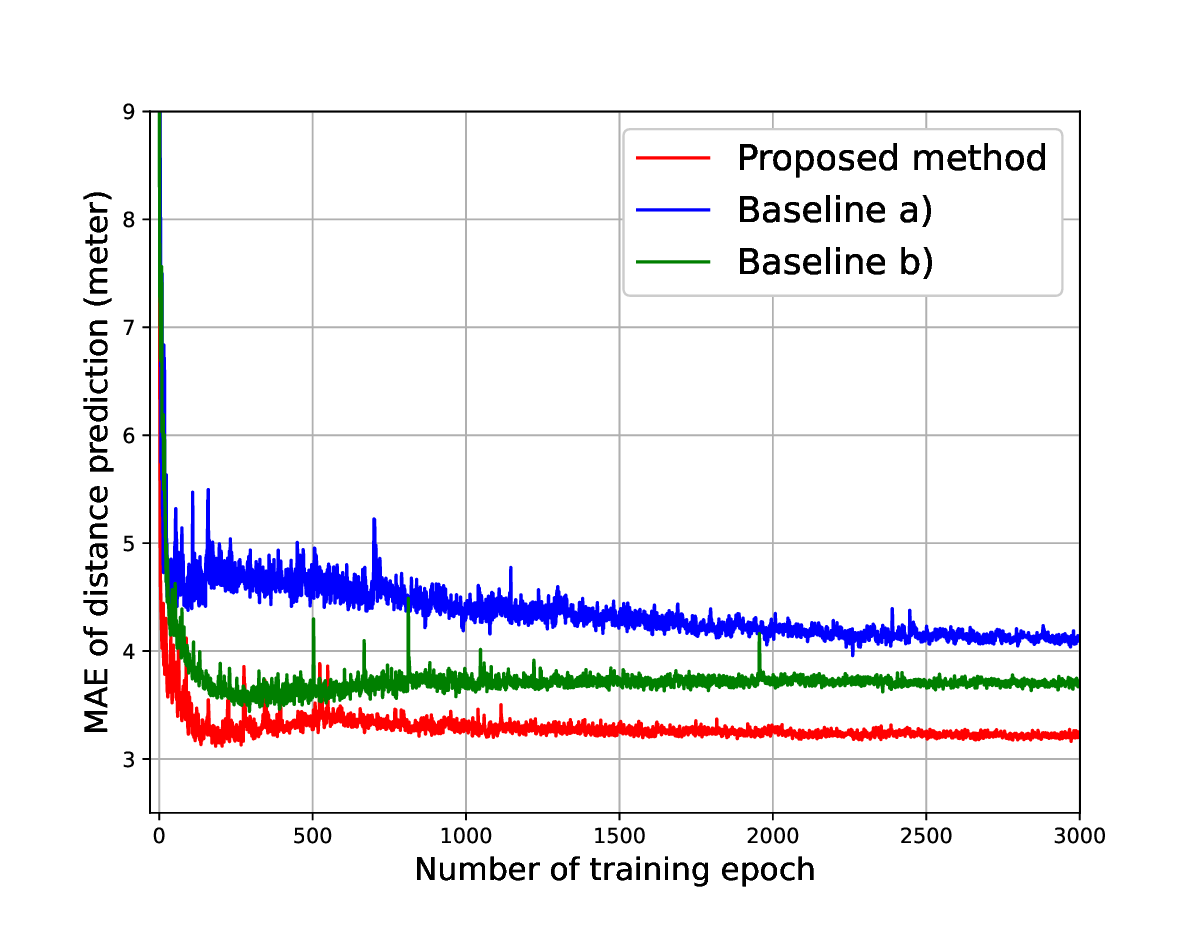}}\vspace{-0.3cm}
\subfigure[Validation positioning errors with $N_{\textrm{T}}=300$.]{\includegraphics[width=8.5cm]{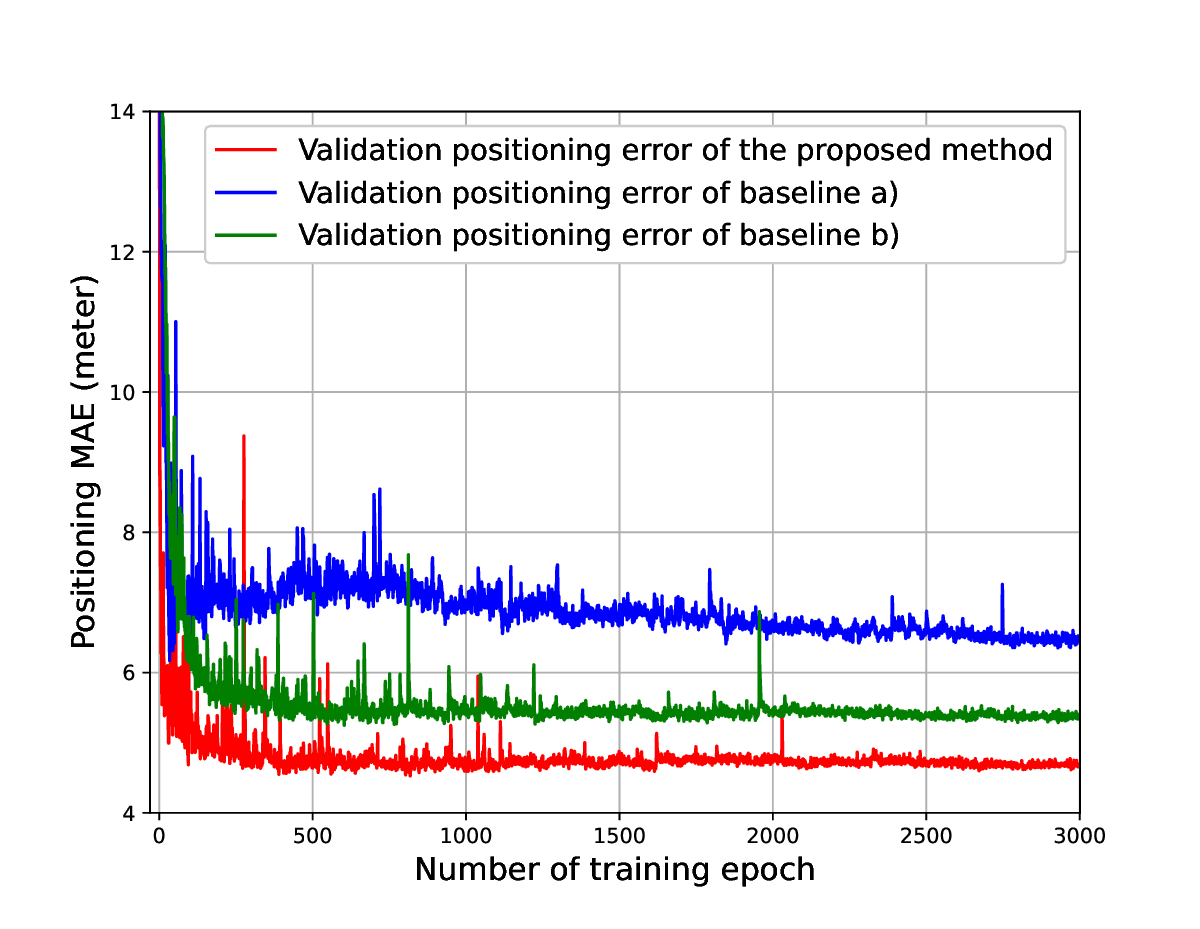}}
\caption{Performance evaluation metrics with $N_{\textrm{T}}=300$.}
\label{fig15}
\vspace{-0.3cm}
\end{figure}

In Fig. \ref{fig15}, we show how the MAEs of the azimuth angle and distance, and positioning errors change as the the number of training epochs increases. From Fig. \ref{fig15}(a), we see that the proposed method reduces the MAE of azimuth angle from $9^{\circ}$ to $6.7^{\circ}$ compared to baseline a), and from $7.5^{\circ}$ to $6.7^{\circ}$ compared to baseline b). This stems from the fact that the proposed method pretrains the positioning model with images and a large number of unlabeled CSI data. Since the purpose of the pretraining of the proposed method is to predict the probability distribution vector of azimuth directions of each vehicle, which is strongly correlated to azimuth angle prediction, the trained model can achieve better accuracy on azimuth angle prediction compared to both baselines a) and b). From Fig. \ref{fig15}(b), we observe that the proposed algorithm can respectively reduce the MAE of distance by up to 22\% and 13\% compared to baselines a) and b), which implies that the pretrained model also has better generalization capacity on vehicle-BS distance prediction despite that the pretraining labels do not explicitly include the distances between vehicles and the BS. This is because the prediction of the probability distribution vector of the vehicle azimuth direction needs to estimate the energy of multipath signals and distinguish the LoS and NLoS links \cite{36}. Since the received signal strength depends on the distance between vehicle and the BS, the trained model is able to extract features related to vehicle distance through pretraining, and thus achieving better generalization capacity on vehicle-BS distance prediction. From Fig. \ref{fig15}(c), we observe that, at the validation dataset, the proposed method can reduce the positioning error by up to 27\%, 12\% compared to baselines a), and b). These gains stem from the fact that the proposed method can effectively avoid overfitting since the proposed method can use pretraining to extract more useful features for positioning.

\begin{figure}[t]
\centering
\includegraphics[width=8.8cm]{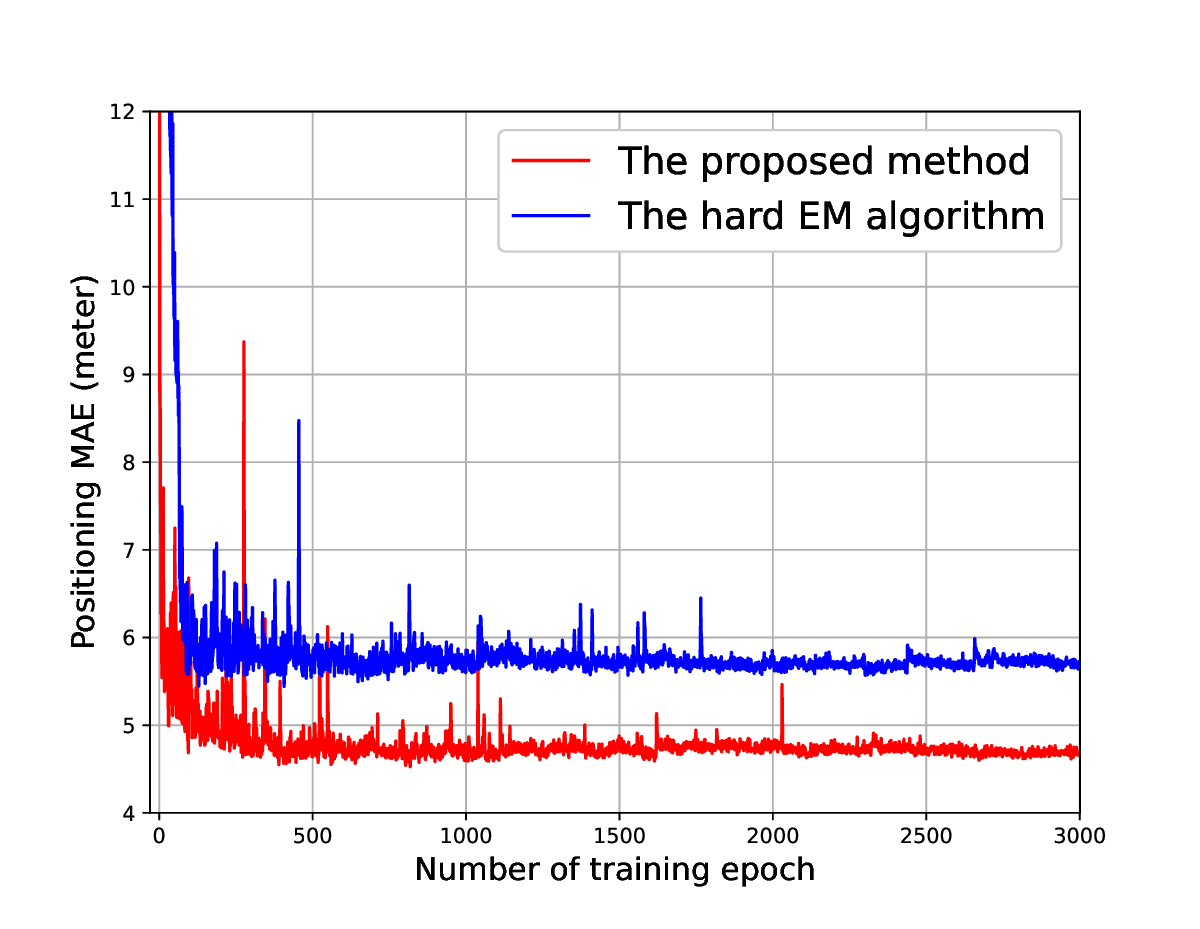}
\centering
\vspace{-0.6cm}
\caption{Positioning errors of the proposed and the hard EM algorithm.}
\label{fig17}
\end{figure}

In Fig. \ref{fig17}, we compared the proposed method to a hard expectation maximization (EM) \cite{MIML,hard_EM} based pretraining method, where the encoder is pretrained by using the Hungarian MSE loss \cite{hungarian_loss} between the predicted azimuth angles of unlabeled CSI and the vehicle azimuth angles obtained from images. From this figure, we see that the proposed method reduces the positioning error by up to 18\% compared to the hard EM algorithm. This stems from the fact that the pretraining objective of the proposed method is to roughly predict the direction of each vehicle, while the objective of the hard EM algorithm is to rigorously minimize the Hungarian MSE loss of azimuth angles. Therefore, compared to our method, the encoder pretrained with the hard EM algorithm extract only features related to angles from CSI, rather than general features related to vehicle locations.

\begin{figure}[t]
\centering
\includegraphics[width=9cm]{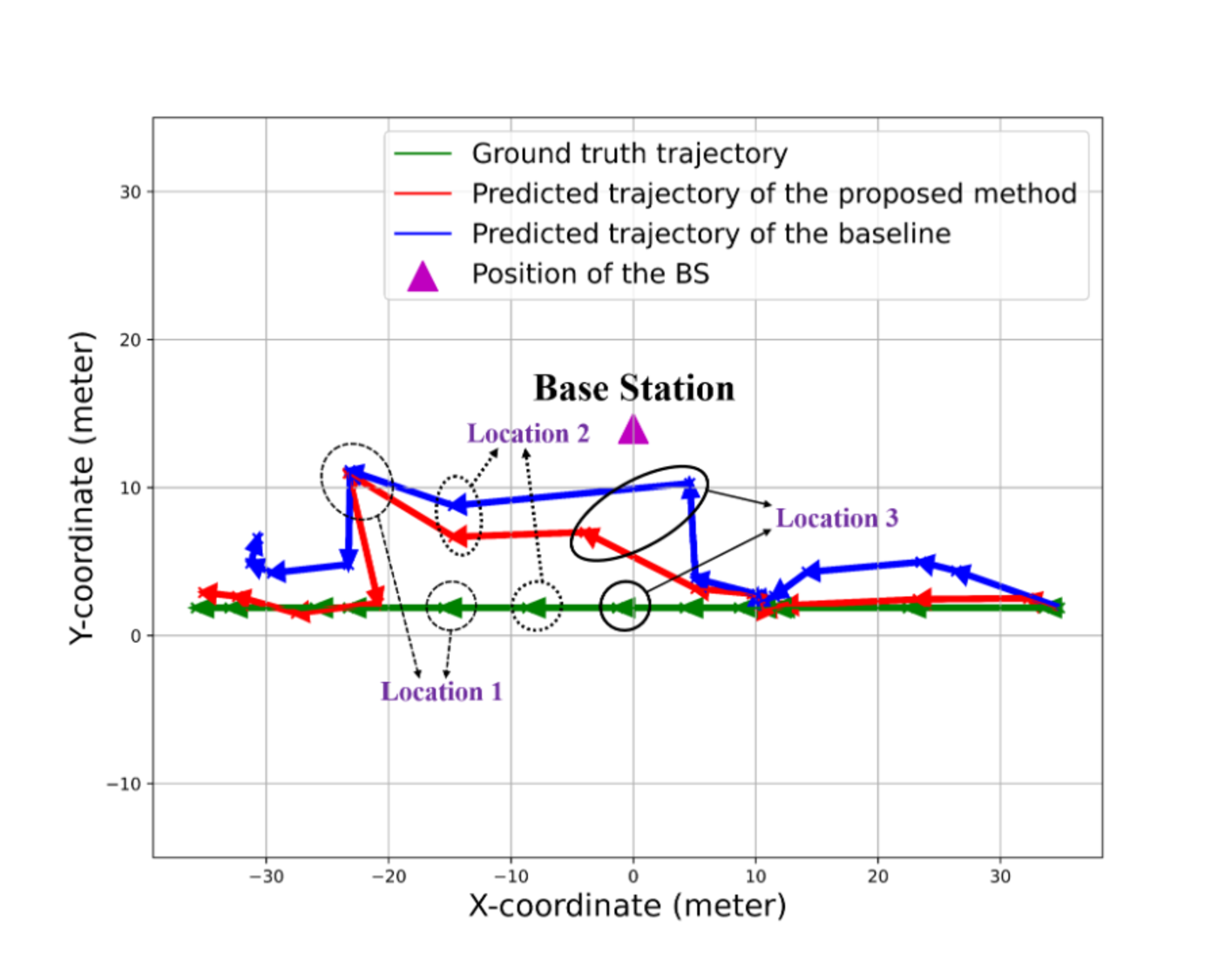}
\centering
\vspace{-0.9cm}
\caption{An example of vehicle positioning with $N_{\textrm{T}}=300$.}
\label{fig16}
\end{figure}

Fig. \ref{fig16} shows an example of vehicle positioning. In this simulation, we randomly select 15 different points on the trajectory of a vehicle, and predict the coordinates of these locations using the proposed method and the baseline a) respectively. From this figure, we see that the location coordinates estimated by the proposed method is closer to the ground truth trajectory compared to the baseline a), which implies that the proposed method can effectively improve the positioning accuracy when the amount of labeled training data is small. From Fig. \ref{fig16}, we also see that the positioning accuracy of three NLoS locations is not high. This is because the localization of vehicles with NLoS links is more complex than localizing LoS vehicles. Thus, to achieve high accuracy for NLoS vehicles, a substantial number of NLoS CSI samples is required. Considering that the majority of the CSI samples in the used dataset is collected under LoS conditions, the number of NLoS samples is not adequate to achieve high positioning accuracy. Nonetheless, it can still be seen from this figure that the positioning errors for NLoS vehicles are lower than that of the baseline, which implies that the proposed method is effective for localizing NLoS vehicles.

\section{Conclusion}
 In this paper, we have designed a novel SSL framework that jointly uses a large sized unlabeled dataset that consists of images and unlabeled CSI data and a small sized labeled dataset that consists of CSI data and their corresponding position coordinates to estimate the positions of the vehicles served by the BS. The proposed framework consists of a pretraining stage and a downstream training stage. In the pretraining stage, the images are used to generate labels for unlabeled CSI data and thus pretraining the model. In the downstream training stage, the model will be retrained on the small sized labeled CSI dataset.  Simulation results have shown that our proposed method can achieve higher positioning precision than the baseline with the same labeled dataset.

\bibliographystyle{IEEEbib}
\bibliography{myrefs}


\end{document}